\documentclass[letterpaper,nameyear,ecta,12pt]{article}
\usepackage{amsmath}
\usepackage{pgfplots}
\usepackage{amsthm,amssymb,mathtools}
\usepackage{enumitem}
\usepackage[linesnumbered,ruled,vlined]{algorithm2e}
\usepackage{comment,todonotes}
\usepackage{natbib}
\usepackage[colorlinks,citecolor=blue,linkcolor=blue]{hyperref}
\usepackage{setspace}
\usepackage[margin=1.25in]{geometry}

\theoremstyle{definition}
\newtheorem{assumption}{Assumption}{}
\newtheorem{definition}{Definition}{}
\newtheorem{lemma}{Lemma}{}
\newtheorem{proposition}{Proposition}{}
\newtheorem{theorem}{Theorem}{}
{}
\newtheorem{example}{Example}{}
{}
\newtheorem{remark}{Remark}{}
\newtheorem{claim}{Claim}{}

\title{ Cores of $3$-distant convex games\thanks{We are grateful to Yukihiko Funaki, Toru Hokari, Shin Kishimoto, Yukio Koriyama, Nobuo Matsubayashi, Satoshi Nakada, and  the participants of European Meeting on Game Theory 2026 (SING21) and Asian Game Theory Conference 2026 (AGT2026) 
for their helpful comments and suggestions.  This work was partially supported by JSPS KAKENHI Grant Numbers  23K01312, 26K21170, and 26K04829 and JST ERATO Grant Number JPMJER2301, Japan.}}

\author{Keisuke Bando\thanks{Department of Industrial and Systems Engineering, Faculty of Science and Technology, Keio University, email: bando@ae.keio.ac.jp.} \and Toshiyuki Hirai\thanks{Department of Economics, Hosei University, e-mail: hirai@hosei.ac.jp.} \and Kenzo Imamura\thanks{Department of Economics, University of Tokyo, email: imamurak@e.u-tokyo.ac.jp.} 
\and Ryuhei Mizutani\thanks{Faculty of Engineering, Tokyo University of Agriculture and Technology, email: \mbox{mizutani}@go.tuat.ac.jp}
}

\begin{document}

\maketitle
\onehalfspacing

\begin{abstract}
Convex cooperative games have a nonempty core and admit an efficient
construction of core allocations. However, economically natural local
frictions may violate convexity. We study the core of $3$-distant
convex games, which permit violations of convexity only between
closely related coalitions. Although the core of such a game may be
empty, we establish a simple necessary and sufficient condition for
its nonemptiness. We also develop a strongly polynomial-time combinatorial
algorithm for finding a core allocation whenever one exists. Finally, we demonstrate
the applicability of the framework to oligopolistic competition with merger synergies
and  profit allocation of joint production with complementary inputs.
\end{abstract}


\section{Introduction}\label{Sect:introduction}

The core is one of the most fundamental solution concepts in cooperative game theory.
It consists of payoff allocations that cannot be improved upon by any coalition.
For transferable utility (TU) games, the concept was originally introduced by \citet{Gillies1959}, 
and a necessary and sufficient condition for the nonemptiness of the core was independently established by \citet{Bondareva1963} and \citet{Shapley1967}. 
The properties of the core have been extensively investigated for several classes of TU games (see, e.g., \citet{PelegSudholter2007CooperativeGames}).

The core exhibits a particularly nice structure in the class of \emph{convex games} introduced by \cite{Shapley1971Cores}: a TU-game $(N,v)$ is said to be convex if 
\begin{align}
\label{ineq:convex}
v(S \cup T) + v(S \cap T) \ge v(S) + v(T)
\end{align}
is satisfied for any pair of coalitions $S$ and $T$.

For convex games, \citet{Shapley1971Cores} shows that the core is nonempty and, in particular, that every marginal worth vector belongs to the core, while \cite{Ichiishi1981} establishes the converse result.
This structural property further implies that 
 a core allocation can be computed efficiently.
Convex games also have wide applicability, as they arise  in several economic settings.\footnote{See, for example, bankruptcy games \citep{ ONeill1982, AumannMaschler1985BankruptcyTalmud, CurielMaschlerTijs1988Bankruptcy}, airport games \citep{littlechild_owen1973, littlechild1974simple}, and  sequencing games \citep{CurielPederzoliTijs1989Sequencing, CurielHamersKlijn2002Survey}.}

However, convexity may fail in the presence of economically natural
local frictions.
The following example
illustrates this point.\footnote{
This example is a special case of an example in \citet{Mizutani2024KDist}; here, we give it a fixed
communication-cost interpretation.
}
\begin{example}\label{ex:fixed_cost}
Let $(N, v)$ be a convex game.
We assume that members of a coalition are
required to coordinate their activities and that forming a coalition
with at least two players incurs a fixed communication cost $c>0$.
By taking this cost into account, we define a new TU game $(N, v')$ by
\[
v'(S) =
\begin{cases}
v(S) & \text{if } |S| \le 1, \\
v(S) - c & \text{if } |S| \ge 2,
\end{cases}
\]
for each $S \subseteq N$.
Then, $(N, v')$  need not be convex.
Indeed,
$$
v'(\{i, j\})  = v(\{i, j\}) - c  <  v(\{i\}) + v(\{j\}) = v'(\{i\}) + v'(\{j\})
$$
for some distinct $i, j \in N$
when $c$ is sufficiently large.
Nevertheless, 
any violation of the convexity condition must involve two distinct singleton coalitions.

\end{example}

Thus, the above example shows that convexity may fail only locally.
Similar phenomena arise in several applications. For example, merger
synergies can generate local violations of convexity 
 in oligopolistic markets (see Subsection \ref{subsec:oligopoly}).

The notion of \emph{3-distant convexity}, originally introduced by
\citet{Mizutani2024KDist} in the context of combinatorial optimization,
provides a natural framework for capturing such local violations.
Intuitively, $3$-distant convexity permits violations of the convexity
condition only between two closely related coalitions that differ
only by replacing one player with another. 
This naturally raises the central question of this paper:
To what extent can the desirable properties of convex
games be preserved under such local violations of convexity?
We first show that these properties are not preserved in full:
The core of a $3$-distant convex game may be empty. 
Nevertheless,
we obtain several positive results.

 First, we provide a simple necessary and sufficient condition for the
nonemptiness of the core (Theorem \ref{thm:ns_core}). 
The condition has a simple interpretation: the sum of the players'
marginal contributions to the grand coalition must be at least the
worth of the grand coalition. Remarkably, whereas the classical Bondareva--Shapley characterization \citep{Bondareva1963,Shapley1967} requires the balancedness inequality
to hold for every balanced collection, our condition requires
checking it for only a single balanced collection. 
 Our proof builds on the classical linear programming characterization
of core nonemptiness underlying the Bondareva--Shapley theorem
\citep{Bondareva1963, Shapley1967}.
A key technical insight is that $3$-distant convexity
imposes strong structural restrictions on dual optimal solutions,
leading to this simple characterization.


Second, we develop a strongly polynomial-time combinatorial algorithm
for finding a core allocation whenever the core is nonempty.
Although our characterization of core nonemptiness is established
through a nonconstructive argument, we provide a separate
constructive proof that directly yields the algorithm. 
 Our algorithm
builds on the standard marginal-vector construction for convex games
but substantially modifies it to accommodate subgames with empty cores.


 We further demonstrate the applicability of our results in economic settings.
First, we analyze an oligopoly game with merger synergies that reduce marginal costs. 
We show that convexity may be violated in the presence of merger synergies while
 $3$-distant convexity is preserved  under a certain synergy level.
Second,  we consider a production game in which
each player allocates her indivisible resource to one of two
complementary inputs for producing a final good, such as iron ore and
coal used in steel production. 
We show that this game is $3$-distant convex even when convexity fails under natural parameter settings.
These examples illustrate that $3$-distant convexity provides a useful framework for analyzing economically relevant situations where convexity does not fully hold.

The rest of this paper is organized as follows. 
The next subsection provides a literature review. 
Section \ref{sec:model} introduces the model. 
Section \ref{sec:main} presents the main theoretical results. 
Section \ref{sec:applications} presents applications of our results. 
Section \ref{sec:extension} concludes. 
All omitted proofs are provided in the Appendix.

\subsection{Related literature}
There are many studies on the core of TU games.
It is well known that a TU game has a nonempty core if and only if it is balanced \citep{Bondareva1963, Shapley1967}.
However, checking the nonemptiness of the core is generally difficult.
Indeed, the problem is NP-complete for certain classes of TU games related to combinatorial optimization problems (see, for example,  \citealp{DengPapadimitriou1994Complexity}; \citealt{DengIbarakiNagamochi1999Core}). 
We provide a new class of TU games for which the nonemptiness of the core can be checked in polynomial time and a core allocation can be found in polynomial time.

\citet{Mizutani2024KDist} considers the minimization problem of $k$-distant concave functions for a fixed positive integer $k (\ge 2)$ and shows that this problem can be solved in polynomial time.
In the proof, \citet{Mizutani2024KDist} provides a polynomial-time algorithm for a more general problem than deciding the nonemptiness of the core and finding a core allocation for $k$-distant convex functions.
By using this algorithm, 
the nonemptiness of the core can be determined in polynomial time, and a core allocation can be found in polynomial time for $3$-distant convex games. The main advantage of our results is that we provide a combinatorial algorithm, whereas the algorithm proposed by \citet{Mizutani2024KDist} is based on the ellipsoid method and is therefore not combinatorial.
Thus, we provide a more efficient way to solve the problem.


Several relaxed notions of convex games have been proposed for TU games, 
including permutational convexity \citep{Granot1982Perm}, 
average convexity \citep{Sprumont1990PMAS, IU1993IJGT}, 
and hyper-additivity \citep{Bahel2021Hyperadditive}. 
 \citet{Driessen1986KConvex} also introduced a class of TU games parameterized by a positive integer $k$, 
called $k$-convex games, which coincide with convex games when $k = n$.
These classes of games are known to have a nonempty core.
The class of $3$-distant convex games is not contained in any of the above classes, 
since a $3$-distant convex game may have an empty core.
 There are also relaxed notions of convexity that do not guarantee a nonempty core.
\citet{DriessenTijs1985Semiconvex} introduced  semi-convexity and showed that every semi-convex
game with at most four players has a nonempty core. In contrast, a
$3$-distant convex game may have an empty core even with only three
players.
 Moreover, \citet{Nunez1998almostconvex} introduced almost convex games.  
They established a necessary and sufficient condition for the nonemptiness of the core of an almost convex game that is strictly stronger than the corresponding condition for $3$-distant convex games, 
although neither class of games contains the other (See Example \ref{ex:empty_core}). 
We apply a new relaxed notion of convexity to TU games and provide several economic applications.


\section{Model}\label{sec:model}

Let $N = \{1, \ldots, n\}$ be a finite set of players.
A nonempty subset of $N$ is called a \emph{coalition}.
A real-valued function $v$ on $2^N$ is called a \emph{characteristic function}, where $v(S)$ denotes the worth of coalition $S \subseteq N$ and $v(\emptyset) = 0$.
The pair $(N, v)$ is called a \emph{transferable-utility} (TU) game.
 Throughout the paper, we assume that 
$v(N) \ge \sum_{i\in N}v(\{i\})$.\footnote{This is a very mild condition as it is weaker than essentiality, that is, $v(N)>\sum _{i\in N}v(\{ i\} )$, which is usually assumed.
 Moreover, the core is empty without this condition.}

A vector $x \in \mathbb{R}^N$ is called an \emph{allocation}.
For each allocation $x$ and $S \subseteq N$, we denote
$x(S) = \sum_{i \in S} x_i$ with $x(\emptyset) = 0$.
The \emph{core} of $(N, v)$ is defined as the set of allocations $x$
such that $x(N) = v(N)$ and
$x(S) \ge v(S)$ for all $S \subseteq N$.
In this case, $x$ is called a \emph{core allocation}.

We say that $(N, v)$ is a \emph{convex game} if 
$v(S \cup T) + v(S \cap T) \ge v(S) + v(T)$ for all $S, T \subseteq N$.
Given any ordering of players $i_1, i_2, \ldots, i_n$,
the corresponding \emph{marginal worth vector} is defined by $$m_{i_\ell} = v(\{i_1, \ldots, i_\ell\}) - v(\{i_1, \ldots, i_{\ell -1}\})
$$ 
for all $\ell = 1, \ldots, n$.
\citet{Shapley1971Cores} shows that every marginal worth vector is in the core
in a convex game.

In this paper, we focus on a relaxed notion of convex games originally introduced by \citet{Mizutani2024KDist}.
\begin{definition}\label{def: 3-convex}
$(N, v)$ is a \emph{3-distant convex game} if 
$v(S \cup T) + v(S \cap T) \ge v(S) + v(T)$ for all $S, T \subseteq N$ with $|S \triangle T| \ge 3$ where $S \triangle T := (S \setminus T) \cup (T \setminus S)$.\footnote{
Indeed, \citet{Mizutani2024KDist} proposes a more general class of \emph{$k$-distant convex games}, 
in which the number $3$ in Definition~\ref{def: 3-convex} is replaced by $k\ge 3$.
}
\end{definition}
 Similarly, we say that a characteristic function $v:2^N\to \mathbb{R}$ is $3$-distant convex if $v$ satisfies the condition in Definition \ref{def: 3-convex}.
In a $3$-distant convex game, the convexity condition may be violated for some pairs 
$S, T \subseteq N$; that is, $v(S \cup T) + v(S \cap T) < v(S) + v(T)$.
Such violations occur only when $|S \setminus T| = 1$ and $|T \setminus S| = 1$.
Therefore, the TU game defined in Example~\ref{ex:fixed_cost} is
$3$-distant convex, because any violation of the convexity condition
can occur only between two distinct singleton coalitions. Moreover,
as we show in Theorem~\ref{thm:ns_core}, its core is nonempty. In
general, however, a $3$-distant convex game may have an empty core.
The following example illustrates this fact.

\begin{example}\label{ex:empty_core}
 Let $(N,v)$ be a three-player game.  
If $(N,v)$ is superadditive, that is, $v(S\cup T)\ge v(S)+v(T)$
for all disjoint coalitions $S,T\subseteq N$, then $(N,v)$ is $3$-distant convex.  Recall that the core of a  three-player superadditive game is nonempty
if and only if 
$v(\{1,2\})+v(\{1,3\})+v(\{2,3\})
\le 2v(N)$.  
Since this inequality does not necessarily hold even under superadditivity, a $3$-distant convex game may have an empty core.
For a $3$-distant convex game with an arbitrary number of players, we will show that a generalization of this condition is a necessary and sufficient condition for the nonemptiness of the core (Theorem \ref{thm:ns_core}).  

The class of three-player games also illustrates the distinction between $3$-distant convex games and almost convex games \citep{Nunez1998almostconvex}.  
A three-player game $(N,v)$ is $3$-distant convex if and only if it is superadditive at $N$, that is, $v(S\cup T)\ge v(S)+v(T)$ for all disjoint coalitions $S,T\subseteq N$ with $S\cup T=N$.  On the other hand, a general TU-game is an almost convex game if the convexity condition \eqref{ineq:convex} holds for any $S,T\subseteq N$ with $S\cup T\neq N$.  Thus, a three-player game $(N,v)$ is almost convex if and only if $v(\{ i\} )+v(\{ j\} )\le v(\{ i,j\} )$ for any distinct $i,j\in N$.  This condition is clearly independent of the condition characterizing $3$-distant convexity for three-player games.

%
\end{example}


We discuss two basic properties of $3$-distant convex games. 
First,  \citet{Mizutani2024KDist} provides an alternative characterization of $3$-distant convexity based on marginal contributions.
It is well known that convexity is characterized by the non-decreasing marginal contribution property.\footnote{
Namely, $v$ is convex if and only if for all $i \in N$ and $S, T \subseteq N \setminus \{i\}$
with $S \subseteq T$,
$v(T \cup \{i\}) - v(T) \ge  v(S \cup \{i\}) -v(S)$.}
For $3$-distant convex games, we have the following result.
\begin{proposition}[\citet{Mizutani2024KDist}]\label{prop:marginal}
$(N, v)$ is $3$-distant convex if and only if for all $i \in N$ and $S, T \subseteq N \setminus \{i\}$ with $S \subseteq T$ and $|T| \ge |S| + 2$,
$v(T \cup \{i\})  - v(T) \ge v(S \cup \{i\}) - v(S)$.\footnote{This result follows from a more general result of $k$-distant convex functions.}
\end{proposition}
Second, the class of $3$-distant convex games is closed under addition. That is, for any $3$-distant convex games $(N, v)$ and $(N, \nu)$,
$v+ \nu$ is also $3$-distant convex.
Therefore, we can restrict attention to the class of $0$-normalized $3$-distant convex games without loss of generality when analyzing the core (see the following remark for more details).
\begin{remark}\label{remark:0-normalization}
A TU game $(N, v)$ is said to be \emph{$0$-normalized} if $v(\{i\}) = 0$ for all $i \in N$. 
Given an arbitrary TU game $(N, \nu)$, define $v(S) = \nu(S) - \sum_{i \in S} \nu(\{i\})$ for all $S \subseteq N$, which is $0$-normalized. 
Then, for any allocation $x \in \mathbb{R}^n$, $x$ is in the core of $(N, \nu)$ if and only if $y$ is in the core of $(N, v)$, where $y_i = x_i - \nu(\{i\})$ for all $i \in N$. 
Moreover, $v$ is $3$-distant convex whenever $\nu$ is $3$-distant convex, since the additive function is convex. 
Therefore, it is without loss of generality to assume $0$-normalization.

\end{remark}

\section{Main results}\label{sec:main}

\subsection{A necessary and sufficient condition }\label{sec:necessary_sufficient}
As illustrated in Example~\ref{ex:empty_core}, the core of a $3$-distant convex game may be empty. 
The following result shows that, for such games, the nonemptiness of the core can be characterized by a simple condition.


\begin{theorem}\label{thm:ns_core}
 Let $(N, v) $ be a $3$-distant convex game. 
Then, the core is nonempty if and only if
\begin{align}
\label{ineq:ns_core}
v(N) \ge \frac{1}{n -1} \sum_{i \in N} v(N \setminus \{i\}).
\end{align}
\end{theorem}

The condition in Theorem~\ref{thm:ns_core} substantially simplifies
the classical necessary and sufficient condition for core
nonemptiness established by \citet{Bondareva1963} and
\citet{Shapley1967}. 
Recall that a collection $\mathcal{B}$ of
nonempty coalitions is \emph{balanced} if there exist positive weights
$(\lambda_S)_{S \in \mathcal{B}}$ such that
$\sum_{S\in\mathcal{B} } \lambda_{S} \chi_S = \mathbf{1}$.\footnote{Note that $\mathbf{1}$ denotes the all-ones vector in $\mathbb{R}^N$, and for each $S\subseteq N$, $\chi _S\in \{ 0,1\} ^N$ denotes the characteristic vector such that the $i$-th coordinate of $\chi _S$ is $1$ if $i\in S$ while it is $0$ otherwise.}
The Bondareva--Shapley theorem states that the core is nonempty if
and only if $\sum_{S\in\mathcal{B}}\lambda_Sv(S)\leq v(N)$ for every balanced collection $\mathcal{B}$ and every associated system of weights $(\lambda_S)_{S \in \mathcal{B}}$.
By contrast, Theorem~\ref{thm:ns_core} shows that, under $3$-distant
convexity, it suffices to check the single balanced collection
consisting of the coalitions $\{ N\setminus \{i\} \mid i \in N \}$  with each coalition assigned weight
$1/(n-1)$. The resulting balancedness inequality is
precisely condition~(\ref{ineq:ns_core}).

Condition~\eqref{ineq:ns_core} admits a simple interpretation: the
sum of the players' marginal contributions to the grand coalition
must be at least the worth of the grand coalition; that is,
\[
\sum_{i\in N}
\bigl(v(N)-v(N\setminus\{i\})\bigr)
\geq v(N).
\]
Thus, remarkably, core nonemptiness is determined entirely by the
players' marginal contributions to the grand coalition.

\citet{Leng2021ADMC} study a class of games distinct from ours,
called almost diminishing marginal contributions (ADMC) games, yet
show that the same condition characterizes core
nonemptiness.\footnote{\citet{Bahel2026Preconvex} study the core of a
broader class of games that contains the class of ADMC games.}
ADMC games require marginal contributions to nonempty coalitions to
be weakly decreasing, whereas $3$-distant convex games require them
to be weakly increasing when the coalition expands by at least two
players. Thus, neither class contains the other.

Despite the coincidence of the core-nonemptiness conditions, the
arguments developed in those studies do not extend to our setting.
 \citet{Leng2021ADMC} show that
the EANSC value, originally introduced by \citet{Moulin1985}
as the equal allocation of nonseparable costs,
\[
x_i
=
v(N)-v(N\setminus\{i\})
+
\frac{ v(N) - 
 \sum_{j\in N}
\left(v(N)-v(N\setminus\{j\} \right) 
}{n} \text{ for all $i \in N$}
\]
belongs to the core for ADMC games satisfying condition (\ref{ineq:ns_core}).
By contrast, the EANSC value need not belong to
the core of a $3$-distant convex game, even when the game is
convex. For example,
consider the three-player game defined by
$v(S)=1$ if $\{1,2\}\subseteq S$ and $v(S)=0$ otherwise. This is a
convex game, but its EANSC value is
$(2/3,2/3,-1/3)$, which does not belong to the core because
$-1/3<v(\{3\})=0$.
Thus, although the same inequality characterizes core nonemptiness,
 the underlying proof is fundamentally different in our setting.


We now show the theorem.
Let $(N,  v)$ be a $3$-distant convex game and  $\nu$ be its $0$-normalization; 
that is, $\nu(S) = v(S)  -\sum_{i \in S} v(\{i\})$ for each $S \subseteq N$.
By Remark \ref{remark:0-normalization}, $\nu$ is $3$-distant convex.
Moreover, $\nu(N) \ge \frac{1}{n-1} \sum_{i \in N} \nu(N \setminus \{i\})$
if and only if $v(N) \ge \frac{1}{n-1} \sum_{i \in N} v(N \setminus \{i\})$.
Therefore, the condition (\ref{ineq:ns_core}) is not affected by $0$-normalization.


In the following, we assume that $(N,v)$ is 0-normalized without any loss of generality.  Thus, $v(N)\ge 0$ by the assumption of this paper.
To prove Theorem~\ref{thm:ns_core}, we use an equivalent variant of
the classical primal-dual linear programming formulation underlying
the Bondareva--Shapley theorem
\citep{Bondareva1963,Shapley1967}.

Let $(N,v)$ be a $0$-normalized $3$-distant convex game with $v(N)\ge 0$.
There exists a core allocation of $(N,v)$ if and only if the optimal value of the following linear program is equal to $v(N)$:
\begin{align*}
(\mathrm{P})\ \ \ \ \ \mathrm{minimize}&\ \ \ x(N)\notag\\
\mathrm{subject\ to}&\ \ \ x(S)\ge v(S)\ \mathrm{for\ every\ }S\subseteq N.\notag
\end{align*}
Consider the dual linear program of (P):
\begin{align*}
(\mathrm{D})\ \ \ \ \ \mathrm{maximize}&\ \ \ \sum_{S\subseteq N}y_S v(S)\\
\mathrm{subject\ to}&\ \ \ y\in \mathbb{R}^{2^N}_{\ge 0},\\&\ \ \ \sum_{S\subseteq N}y_S \chi_S=\mathbf{1},
\end{align*}
 where $\mathbf{1}$ denotes the all-ones vector in $\mathbb{R}^N$, and for each $S\subseteq N$, $\chi _S\in \{ 0,1\} ^N$ denotes the characteristic vector such that the $i$-th coordinate of $\chi _S$ is $1$ if $i\in S$ while it is $0$ otherwise.
Let $y^*$ be an optimal solution to (D) minimizing
\begin{align}
\label{func:uncrossing}
\sum_{S\subseteq N}y_S|S||N\setminus S|.
\end{align}
We denote the support of $y^*$ by $\mathcal{L}=\{S\subseteq N\mid y^*_S>0,S\ne \emptyset\}$.
Since $v$ is $3$-distant convex, the following lemma holds by the uncrossing argument\footnote{
Uncrossing is a standard technique in combinatorial optimization; see,
for example, \citet{Frank2011} and \citet{Schrijver2003}.
The uncrossing argument works as follows. Suppose that there exist $S,T\in \mathcal{L}$ that do not satisfy the two conditions in Lemma \ref{lem:uncrossing}.
Let $\alpha:=\min\{y^*_{S},y^*_{T}\}$. 
Then, we execute the uncrossing operation, that is, decrease $y^*_S$ and $y^*_T$ by $\alpha$ and increase $y^*_{S\cup T}$ and $y^*_{S\cap T}$ by $\alpha$.
By $3$-distant convexity of $v$, the obtained solution is another optimal solution to (D) with the smaller value of (\ref{func:uncrossing}), which contradicts the definition of $y^*$.
} for 3-distant convex functions:
\begin{lemma}[Claim 3.2 in \cite{Mizutani2024KDist}]
\label{lem:uncrossing}
For every pair of sets $S,T\in \mathcal{L}$, one of the following two conditions holds:
\begin{itemize}
    \item  either $S\subseteq T$ or $S\supseteq T$.
    \item $|S\bigtriangleup T|\le 2$.
\end{itemize}
\end{lemma}

Let $\mathcal{C}_N\subseteq 2^N$ be a family defined as follows: 
\begin{align*}
 &\mathcal{C}_N=\{N\setminus \{i\}\mid i\in N\}\cup \{N\}.
\end{align*}
The following lemma establishes strong structural restrictions on $\mathcal{L}$, which plays a key role in the proof of Theorem \ref{thm:ns_core}:

\begin{lemma}
\label{lem:C_N}
$\mathcal{L}\subseteq \mathcal{C}_N$.
\end{lemma}

\begin{proof}
Let $\mathcal{C}_\emptyset=\{\{i\}\mid i\in N\}\cup \{\emptyset\}$.
We first show $\mathcal{L}\subseteq \mathcal{C}_\emptyset \cup \mathcal{C}_N$.
Suppose to the contrary that $\mathcal{L}\not\subseteq \mathcal{C}_\emptyset \cup \mathcal{C}_N$.
Take $X\in \mathcal{L}\setminus (\mathcal{C}_\emptyset \cup \mathcal{C}_N)$.
Since $X\notin \mathcal{C}_\emptyset \cup \mathcal{C}_N$, we have $2\le |X|\le n-2$.
Combined with Lemma \ref{lem:uncrossing}, this implies that for each $S\in \mathcal{L}$, we have either $|S\setminus X|=0$, $|X\setminus S|=0$, or $|S\bigtriangleup X|\le 2\le |X|$ and $|X|\le n-2\le |N\setminus (S\bigtriangleup X)|$.
The last case implies
\begin{align*}
|S\setminus X|=|(S\bigtriangleup X)\setminus (X\setminus S)|\le |X\setminus (X\setminus S)|=|S\cap X|
\end{align*}
and
\begin{align*}
|X\setminus S|=|X\setminus (S\cap X)|\le |N\setminus ((S\bigtriangleup X)\cup (S\cap X))|=|N\setminus (S\cup X)|.
\end{align*}
Then, in all cases, we have
\begin{align}
\label{ineq:xsratio}
\dfrac{|S\cap X|}{|X|}\ge \dfrac{|S\setminus X|}{|N\setminus X|}
\end{align}
for every $S\in \mathcal{L}$.
Indeed, if $|S\setminus X|=0$, then the right hand side of (\ref{ineq:xsratio}) is equal to $0$, hence (\ref{ineq:xsratio}) holds. 
If $|X\setminus S|=0$, then the left hand side of (\ref{ineq:xsratio}) is equal to $1$, hence it holds.
If $|S\setminus X|\le |S\cap X|$ and $|X\setminus S|\le |N\setminus (S\cup X)|$, then we have
\begin{align*}
&|S\cap X|\cdot |N\setminus X|-|S\setminus X|\cdot |X|\\&=|S\cap X|\cdot (|N|-|X|)-|S\setminus X|\cdot |X|\\&=|S\cap X|\cdot |N|-|S|\cdot |X|\\&=|S\cap X|\cdot |N\setminus (S\cup X)|+|S\cap X|\cdot |S\cup X|-|S|\cdot |X|\\&=|S\cap X|\cdot |N\setminus (S\cup X)|+|S\cap X|\cdot (|S|+|X\setminus S|)-|S|\cdot (|S\cap X|+|X\setminus S|)\\&=|S\cap X|\cdot |N\setminus (S\cup X)|-|S\setminus X|\cdot |X\setminus S|\\&\ge |S\setminus X|\cdot |X\setminus S|-|S\setminus X|\cdot |X\setminus S|=0.
\end{align*}
This implies the inequality (\ref{ineq:xsratio}).

Since $y^*$ is a feasible solution to (D), we have
\begin{align}
\label{eq:feasible}
\sum_{S\in \mathcal{L}}y^*_S\chi_S=\mathbf{1}.
\end{align}
Combined with $X\in \mathcal{L}$, this implies
\begin{align}
\label{eq:oneminusx}
\sum_{S\in \mathcal{L}\setminus \{X\}}y^*_S\chi_S=\mathbf{1}-y^*_X\chi_X.
\end{align}
Hence, summing the $i$-th coordinates of both sides of (\ref{eq:oneminusx})  over all $i\in X$, we obtain
\begin{align}
\label{eq:sumx}
\sum_{S\in \mathcal{L}\setminus \{X\}}|S\cap X|\cdot y^*_S=|X|\cdot (1-y^*_X).
\end{align}
Similarly, summing the $i$-th coordinates of both sides of (\ref{eq:oneminusx})  over all $i\in N\setminus X$, we obtain
\begin{align}
\label{eq:sumnx}
\sum_{S\in \mathcal{L}\setminus \{X\}}|S\setminus X|\cdot y^*_S=|N\setminus X|.
\end{align}
Therefore, by (\ref{eq:sumx}) and (\ref{eq:sumnx}) we have
\begin{align}
\label{ineq:ratio}
\sum_{S\in \mathcal{L}\setminus \{X\}}\dfrac{|S\cap X|}{|X|}\cdot y^*_S=1-y^*_X<1=\sum_{S\in \mathcal{L}\setminus \{X\}}\dfrac{|S\setminus X|}{|N\setminus X|}\cdot y^*_S.
\end{align}
This contradicts the inequality (\ref{ineq:xsratio}).
Thus, we have $\mathcal{L}\subseteq \mathcal{C}_\emptyset \cup \mathcal{C}_N$.



We now show $\mathcal{L}\subseteq\mathcal{C}_N$.
If $|N|=2$, then we have $\mathcal{C}_N=2^N\setminus \{\emptyset\}$.
Since $\emptyset\notin \mathcal{L}$, this implies $\mathcal{L}\subseteq \mathcal{C}_N$, as desired.
Assume that $|N|\ge 3$.
Suppose to the contrary that $\mathcal{L}\not\subseteq \mathcal{C}_N$.
Then, since $\mathcal{L}\not\subseteq \mathcal{C}_N$ and $\mathcal{L}\subseteq \mathcal{C}_\emptyset \cup \mathcal{C}_N$ by the above argument, it follows that $\mathcal{L}\cap \mathcal{C}_\emptyset$ is nonempty.
Take $\{i\}\in \mathcal{L}\cap \mathcal{C}_\emptyset$ maximizing the value of $y^*_{\{i\}}$.

Consider the case when there exists $j\in N$ such that $\{j\}\notin \mathcal{L}\cap \mathcal{C}_\emptyset$, or $\{j\}\in \mathcal{L}\cap \mathcal{C}_\emptyset$ and $y^*_{\{i\}}>y^*_{\{j\}}$.
Then, by (\ref{eq:feasible}) there exists $X\in \mathcal{L}\cap \mathcal{C}_N$ such that $i\notin X$ and $j\in X$.
Since $X\in \mathcal{C}_N$, we have $X=N\setminus \{i\}$.
Let $\alpha=\min\{y^*_{\{i\}},y^*_{N\setminus \{i\}}\}$.
We define $y'$ as the vector obtained from $y^*$ by decreasing $y^*_{\{i\}}$ and $y^*_{N\setminus \{i\}}$ by $\alpha$, and increasing $y^*_{N}$ by $\alpha$.
Then, $y'$ is a feasible solution to (D).
Since $|N|\ge 3$ and $v$ is 3-distant convex, we have $v({\{i\}})+v(N\setminus \{i\})\le v(N)$, which implies
\begin{align*}
\sum_{S\subseteq N}y'_Sv(S)\ge \sum_{S\subseteq N}y^*_Sv(S).
\end{align*}
Hence, $y'$ is an optimal solution to (D).
We also have
\begin{align*}
\sum_{S\subseteq N}y'_S|S||N\setminus S|<\sum_{S\subseteq N}y^*_S|S||N\setminus S|.
\end{align*}
This contradicts the definition of $y^*$.

Consider the case when for all $j\in N$, $\{j\}\in \mathcal{L}\cap \mathcal{C}_\emptyset$ and $y^*_{\{i\}}=y^*_{\{j\}}$.
Let $\alpha=y^*_{\{i\}}$.
We define $y'$ as the vector obtained from $y^*$ by decreasing $y^*_{\{j\}}$ by $\alpha$ for all $j\in N$, and increasing $y^*_N$ by $\alpha$.
Then, $y'$ is a feasible solution to (D).
Since $(N,v)$ is 0-normalized and $v(N)\ge 0$,
we have $\sum_{j\in N}v(\{j\})=0\le v(N)$, which implies
\begin{align*}
\sum_{S\subseteq N}y'_Sv(S)\ge \sum_{S\subseteq N}y^*_Sv(S).
\end{align*}
Hence, $y'$ is an optimal solution to (D).
We also have
\begin{align*}
\sum_{S\subseteq N}y'_S|S||N\setminus S|<\sum_{S\subseteq N}y^*_S|S||N\setminus S|.
\end{align*}
This contradicts the definition of $y^*$.
\end{proof}

\noindent
{\bf Proof of Theorem \ref{thm:ns_core}.}
There exists a core allocation of the 3-distant convex game $(N,v)$ if and only if the optimal value of (D) is equal to $v(N)$.
Hence, it suffices to show that the optimal value of (D) is equal to $v(N)$ if and only if (\ref{ineq:ns_core}) holds.
We first show the necessity part.
Let $y\in \mathbb{R}_{\ge 0}^{2^N}$ be a vector defined by
\begin{align*}
y_S=\begin{cases}
    \dfrac{1}{n-1} & \mathrm{if\ }S=N\setminus \{i\}\ \mathrm{for\ some\ }i\in N,\vspace{4pt}\\ 
    \ \ \ 0 & \mathrm{otherwise}.
\end{cases}
\end{align*}
Then, $y$ is a feasible solution to (D).
Moreover, by the assumption, the optimal value of (D) is $v(N)$.
Hence, the inequality (\ref{ineq:ns_core}) holds.

We next show the sufficiency part.
By Lemma \ref{lem:C_N}, there exists an optimal solution $y^*$ to (D) such that its support $\mathcal{L}=\{S\subseteq N\mid y^*_S>0,S\ne \emptyset\}$ is included in $\mathcal{C}_N$.
Since $y^*$ is a feasible solution to (D), we have
\begin{align*}
\sum_{S\in \mathcal{C}_N}y^*_S \chi_S=\mathbf{1}.
\end{align*}
This implies
\begin{align*}
\sum_{S\in \mathcal{C}_N\setminus \{N\}}y^*_S \chi_S=\mathbf{1}-y^*_N\cdot \mathbf{1}=(1-y^*_N)\cdot \mathbf{1}.
\end{align*}
Since $\mathcal{C}_N\setminus \{N\}=\{N\setminus \{i\}\mid i\in N\}$, we have
\begin{align*}
\sum_{i\in N}y^*_{N\setminus \{i\}} \chi_{N\setminus \{i\}}=(1-y^*_N)\cdot \mathbf{1}.
\end{align*}
Hence, we have
\begin{align*}
\sum_{i\in N}y^*_{N\setminus \{i\}} \chi_{\{i\}}=\sum_{i\in N}y^*_{N\setminus \{i\}}\cdot \mathbf{1}-\sum_{i\in N}y^*_{N\setminus \{i\}} \chi_{N\setminus \{i\}}=\left(\sum_{i\in N}y^*_{N\setminus \{i\}}-1+y^*_N\right)\cdot \mathbf{1}.
\end{align*}
This implies
\begin{align*}
y^*_{N\setminus \{j\}}=\sum_{i\in N}y^*_{N\setminus \{i\}}-1+y^*_N
\end{align*}
for all $j\in N$.
Therefore, we have
\begin{align*}
\sum_{i\in N}y^*_{N\setminus \{i\}}=n\cdot \left(\sum_{i\in N}y^*_{N\setminus \{i\}}-1+y^*_N\right).
\end{align*}
This implies
\begin{align*}
\sum_{i\in N}y^*_{N\setminus \{i\}}=\dfrac{n}{n-1}\cdot (1-y^*_N)
\end{align*}
and
\begin{align*}
y^*_{N\setminus \{i\}}=\dfrac{1}{n-1}\cdot (1-y^*_N)
\end{align*}
for all $i\in N$.
Thus, we have
\begin{align*}
&\sum_{S\subseteq N}y^*_Sv(S)=\sum_{S\in \mathcal{C}_N}y^*_Sv(S)=y^*_Nv(N)+(1-y^*_N)\cdot \dfrac{1}{n-1}\sum_{i\in N}v(N\setminus \{i\})\\&\le y^*_Nv(N)+(1-y^*_N)v(N)=v(N).
\end{align*}
The inequality follows from (\ref{ineq:ns_core}).
Therefore, the optimal value of (D) is at most $v(N)$.
Since $y\in \mathbb{R}^{2^N}_{\ge 0}$ defined by $y_N=1$ and $y_S=0$ for all $S\subsetneq N$ is a feasible solution to (D), the optimal value of (D) is at least $v(N)$.
Thus, the optimal value of (D) is $v(N)$.
This completes the proof.

\subsection{Finding a core allocation}\label{subsec:algorithm}

Theorem~\ref{thm:ns_core} provides a simple test for core
nonemptiness, but it does not by itself identify a core allocation.
For convex games, a core allocation can be obtained immediately by computing
a marginal worth vector. This approach is no longer
available for $3$-distant convex games: even when the core is
nonempty, it may contain no marginal worth vector.\footnote{
For example, consider the three-player game defined by
$v({i})=0$, $v({i,j})=2$ for all distinct $i,j\in N$, and
$v(N)=3$. This game is superadditive and hence $3$-distant convex.
Its core consists of the unique allocation $(1,1,1)$, whereas every
marginal worth vector is a permutation of $(0,2,1)$ and lies outside
the core.}
Moreover, as discussed in
Section~\ref{sec:necessary_sufficient}, the EANSC value, which yields
a core allocation for ADMC games satisfying (\ref{ineq:ns_core}),  
need not belong
to the core in our setting. Thus, core nonemptiness does not
immediately provide a canonical way to construct a core allocation.

We provide a combinatorial polynomial-time algorithm
to find a core allocation whenever the core is nonempty. 
To this end, we provide a constructive proof of Theorem \ref{thm:ns_core}.
Building on this proof, we present our algorithm.
This algorithm extends the standard greedy algorithm in the sense that it returns a marginal worth vector when the game is convex. 

\subsubsection*{A constructive proof  of Theorem \ref{thm:ns_core}}

In the following, we show Theorem \ref{thm:ns_core} by a mathematical induction on $|N|$.  The proofs of lemmas are relegated to Appendix A.
Suppose that $|N| = 2$.
Note that (\ref{ineq:ns_core}) is equivalent to $v(N) \ge 0$
since $v$ is $0$-nomalized.
Then, $(x_1, x_2) = (0, v(N))$ is a core allocation.
Let $n \ge 3$. 
We assume that 
the core is nonempty for any $0$-nomalized and $3$-distant convex game $(N, v)$ satisfying (\ref{ineq:ns_core}) and $v(N) \ge 0$
such that $|N| \le n-1$
(Induction hypothesis).

We consider a $0$-nomalized and $3$-distant convex game $(N, v)$ with $v(N) \ge 0$ where $|N| = n$.
Pick any player $\hat{i} \in N$.
Let $(N \setminus \{ \hat{i} \}, v_{-\hat{i}})$ be the subgame obtained by removing player $\hat{i}$, 
where $v_{-\hat{i}}(S) = v(S)$ for all $S \subseteq N \setminus \{ \hat{i} \}$.
The main difficulty is that the nonemptiness of the core is not guaranteed for the subgame, though $(N\setminus \{ \hat i\} ,v_{-\hat i})$ is $3$-distant convex. 
Specifically, there are three possibilities.

\begin{itemize}
\item Case 1.  $v(N \setminus \{\hat{i}\}) \ge 0$ and  $(N \setminus \{ \hat{i} \}, v_{-\hat{i}})$ satisfies (\ref{ineq:ns_core}).
\item Case 2.  $v(N \setminus \{\hat{i}\}) \ge 0$ and 
$(N \setminus \{ \hat{i} \}, v_{-\hat{i}})$ does not satisfy (\ref{ineq:ns_core}).
\item Case 3.  $v(N \setminus \{\hat{i}\}) < 0$.
\end{itemize}

In Example \ref{ex:fixed_cost}, Case 3 may happen.
Case 2 also happens even if the original game satisfies (\ref{ineq:ns_core}).
The following example illustrates this fact.
\begin{example}
Let $N = \{1, 2, 3, 4\}$.
Let $v$ be 
\begin{align*}
v(S) = \begin{cases}
 6  & \text{if } S = N, \\
 3  & \text{if } |S| = 2, 3, \\
 0  & \text{if } |S| \le 1.\\
\end{cases}
\end{align*}
Then, $(N,v)$ is a 3-distant convex game. Moreover,
$(N,v)$ satisfies (\ref{ineq:ns_core}) 
by $v(N) = 6 > \frac{1}{3} \sum_{i \in N} v(N \setminus \{i\})  =  4$
while  $(N\setminus \{ i\} ,v_{-i})$ does not by
$v(N \setminus \{i\}) = 3  < \frac{1}{2} \sum_{j \in N \setminus \{i\} } v(N \setminus \{i, j\})  =  4.5$ for all $i \in N$.
Note that the core of $(N \setminus \{ i \}, v_{-i})$ is empty for all $i \in N$.
\end{example}
We construct a core allocation of $(N, v)$ for each case as follows.
We first consider Case 1; that is, $v(N \setminus \{\hat{i}\}) \ge 0$ and  $(N \setminus \{ \hat{i} \}, v_{-\hat{i}})$ satisfies (\ref{ineq:ns_core}).
Thus, the core of $(N\setminus \{ \hat i\} ,v_{-\hat i})$ is nonempty by induction hypothesis.
Let  $(x_{i})_{i \in N \setminus \{ \hat{i} \}}$  be a core allocation for $(N \setminus \{ \hat{i} \}, v_{-\hat{i}})$.  
Let $x_{\hat{i}} = v(N) - v(N \setminus \{\hat{i}\})$
and $x = (x_i)_{i \in N}$.
The following lemma shows that we can determine whether \(x\) is a core allocation of the original game by checking only \(|N|-1\) equations.

\begin{lemma}\label{claim:core0}
If $x(S) \ge v(S)$ for all $S \subseteq N$ such that $\hat{i} \in S$ and $|S| = |N| -1$,
then $x$ is a core allocation for $(N, v)$.
\end{lemma}

When $x$ is a core allocation for $(N,v)$,  the proof is done.
In the following, we assume that $x$ is not a core allocation.
Let $x^*_i = v(N)- v(N \setminus \{i\})$ for all $i \in N$.
Let $N^+ = \{ i \in N \mid x_i > x^*_i\} $, $N^{-} = \{ i \in N \mid x^{*}_i > x_i\} $, and $N^{=} =  \{ i \in N \mid x_i = x^{*}_i\}$.
Note that $\hat{i} \in N^{=}$.
The following property is crucial.
\begin{lemma}\label{claim:core1}
\begin{itemize}
\item[(a)]  $N^+ \neq \emptyset$ and $N^{-} \neq \emptyset$.
\item[(b)] $ x(N^+) - x^{*}(N^{+} ) \le x^*(N^{-}) - x(N^{-}) $.
\end{itemize}
\end{lemma}

Let members in $N^+$ transfer $x(N^+) - x^{*}(N^{+} ) > 0$
to members in $N^{-}$. 
By Lemma \ref{claim:core1}, we can construct an allocation $x'$ 
 satisfying the following two conditions:\footnote{
For example, $x'$ is constructed by
the following procedure:
\begin{itemize}
\item Fix an ordering of $N^{-}$ and write $N^{-} = \{i_1, \dots, i_m\}$.
\item Let $E =
x(N^+) - x^{*}(N^{+})$ and   $d_{k} = x^*_{i_k} - x_{i_k}$ for all $k = 1, \ldots, m$.
\item  Let $r_{i_1} = \min \{d_1, E\}$ and  $r_{i_k} = \min \{d_k, E - \sum^{k-1}_{k' = 1} r_{i_{k'}} \}$ for all $k = 2, \ldots, m $.
\item Define $x'$ by 
\begin{align*}
x'_{i} = \begin{cases}
x_i + r_i &\text{ if } i \in N^- \\
x^*_i   & \text{ otherwise.} \\
\end{cases}
\end{align*}
\end{itemize}}
\begin{align}
&x'_i = x^*_i \text{ for all } i \in N^+ \cup N^{=} \text{ and } 
x_i \le x'_i \le  x^{*}_i   \text{ for all } i \in  N^-,\label{eq:algorithm0}\\ 
&x(N^{+})  - x^{*}(N^{+}) =  x'(N^{-}) - x(N^{-}).\label{eq:algorithm1}
\end{align}
The following claim completes the proof for Case 1.
\begin{lemma}\label{claim1:case1_core}
$x'$ is a core allocation.
\end{lemma}

We next consider Case 2; that is, $v(N \setminus \{\hat{i}\}) \ge 0$ and 
$(N \setminus \{ \hat{i} \}, v_{-\hat{i}})$ does not satisfy (\ref{ineq:ns_core}).
Then, 
we have 
\begin{align*}
\epsilon := \frac{ \sum_{i \in N \setminus \{\hat{i}\} } v(N \setminus \{\hat{i}, i\}) }{n-2} -  v(N \setminus \{ \hat{i} \} )  > 0. 
\end{align*}
 Note that $\epsilon $ is well-defined by $n\ge 3$.  
Define the allocation $y$ by
\begin{align}
y_i =
\begin{cases}
v(N) - v(N \setminus \{ \hat{i} \}) - \epsilon & \text{ if } i = \hat{i}. \\
v(N \setminus \{ \hat{i} \})  - v(N \setminus \{ \hat{i}, i \}) + \epsilon & \text{ if } i \in N \setminus \{ \hat{i} \}.\label{eq:algorithm2}
\end{cases}
\end{align}
The following claim completes the proof
for Case 2.
\begin{lemma}\label{claim:imputation}
$y$ is a core allocation for $(N, v)$. 
\end{lemma}

We finally consider Case 3; that is, $v(N \setminus \{\hat{i}\}) < 0$.
The following claim completes the proof.
\begin{lemma}\label{claim:imputation_minus}
Let $z_{\hat{i} } = v(N)$ and $z_i = 0$ for any $i \in N \setminus \{ \hat{i} \}$. Then,
$z$ is a core allocation for $(N, v)$.
\end{lemma}

\subsubsection*{An algorithm}
The constructive proof directly implies an algorithm defined as follows.
Let $(N, v)$ be a $0$-normalized and 3-distant convex game that satisfies (\ref{ineq:ns_core}) and $v(N) \ge0$.
For each nonempty $N' \subseteq N$, 
we simply write
$(N', v)$ for the subgame of $(N, v)$ by restricting the player set to $N'$.
Fix an ordering of $N$ and write $N = \{1, \dots, n\}$.
We define a function $f$  on $\{2, \ldots, n\}$ recursively as follows.
Here, $f(\ell)$ is $0$ or an element in $\mathbb{R}^{\ell}$
where $f(\ell) = 0$ means that the core is empty for $(\{1, \ldots, \ell\}, v)$
and $f(\ell) \neq  0$ means that $f(\ell)$ is a core allocation  for
 $(\{1, \ldots, \ell\}, v)$.
Let $f(2 ) = (0, v(\{1, 2\}))$.
Suppose that $f(\ell -1)$ is defined for $\ell-1 =2, \dots, n-1 $.

\begin{itemize}

\item Step $0$:  
Check whether $(\{1, \ldots, \ell \}, v)$ satisfies (\ref{ineq:ns_core}). 
If it does not satisfy (\ref{ineq:ns_core}), set $f(\ell ) = 0$.
Suppose that  $(\{1, \ldots, \ell \}, v)$  satisfies (\ref{ineq:ns_core}).
There are three possibilities corresponding to Cases 1, 2, and 3.
\begin{itemize}
\item[(1)] If $v(\{1, \ldots, \ell-1\}) \ge 0$ and $f(\ell - 1) \neq 0$, then go to Step $1$.
\item[(2)]If $v(\{1, \ldots, \ell-1\}) \ge0$ and $f(\ell - 1) = 0$, then go to Step 2.
\item[(3)] If $v(\{1, \ldots, \ell-1\}) < 0$, then go to Step 3.
\end{itemize}
\item Step $1$:  Let $x = (f(\ell-1), v(\{1, \ldots, \ell \}) - v(\{1, \ldots, \ell-1\}))$.
If $x(S) \ge v(S)$ for all $S \subseteq \{1, \ldots, \ell\}$ with $\ell \in S$ and $|S| =\ell - 1$,
then set $f(\ell)= x$.
Otherwise, construct $x'$ satisfying (\ref{eq:algorithm0}) and (\ref{eq:algorithm1}), and set $f(\ell)= x'$.

\item Step $2$:  Let $y$ be an allocation defined in (\ref{eq:algorithm2}).
Set $f(\ell) = y$.

\item Step $3$: Let $z$ be an allocation defined in Lemma \ref{claim:imputation_minus}.
Set $f(\ell) = z$.
\end{itemize}
The proof argument shows $f(n)$ is a core allocation.
Moreover, $f(n)$ is constructed in  polynomial time $O(n^2)$ since each iteration $\ell=3,\ldots,n$ of the above algorithm requires $O(n)$ evaluations of the characteristic function and $O(n)$ arithmetic operations.

\begin{theorem}\label{thm:algorithm}
For any $0$-normalized and 3-distant convex game that satisfies $v(N) \ge 0$ and \eqref{ineq:ns_core}, 
$f(n)$ is a core allocation.
\end{theorem}

\section{Applications}\label{sec:applications}

\subsection{ Bertrand Oligopoly with  Synergy effects }\label{subsec:oligopoly}

 We consider a TU game arising from a differentiated-products Bertrand
oligopoly, a setting in which firms' merger incentives have been studied by
\citet{DeneckereDavidson1985}. We allow mergers to generate cost-reduction synergies, as considered in the horizontal-merger literature \citep{FarrellShapiro1990}. In particular, as in \citet{Williamson1968}, we represent such synergies as a merger-induced reduction in constant unit costs.
Our model generalizes the model of \citet{Lardon2020Convexity} by incorporating synergy effects.
In the absence of synergy effects, \citet{Lardon2020Convexity} shows that the game is convex.
We show that convexity may fail in the presence of synergy effects,
while $3$-distant convexity is satisfied under certain levels of synergy.
We provide a sufficient condition on the level of synergy effects that guarantees $3$-distant convexity of the game.
As a by-product, we also provide a sufficient condition under which the game remains convex.\footnote{The convexity of oligopoly market-related games has been extensively studied in the literature; see, for example, \citet{Meinhardt1999CommonPool}, \citet{Zhao1999ConvexityOligopoly}, \citet{DriessenMeinhardt2001}, \citet{Lardon2019Core}, and \citet{HouLardonDriessen2025Bertrand}.}

Let $N = \{1,\ldots, n\}$ be the set of firms.
Each firm has an identical linear cost function
where  $c > 0$ is the marginal cost.
For each price vector $(p_1, \ldots, p_n) \in \mathbb{R}_{+}^n$, the profit of $i \in N$ is given by
$$
\pi_i (p_1, \ldots, p_n) = D_i(p_1, \ldots, p_n) (p_i-c),
$$
where  the demand function of firm $i$, proposed by \citet{Shubik1980MarketStructure}, is given by
$$
D_i(p_1, \ldots, p_n) = V - p_i - r\left(p_i -\frac{ \sum_{j \in N}p_ j}{n} \right),
$$
where $V > 0$ is the intercept of the demand curve and $r>0$ is
the substitutability parameter.
Note that a larger $r$ implies that products are less differentiated.

We consider the situation where the firm merger involves synergy effects that reduce the marginal cost.
Specifically, for each nonempty $S \subseteq N$ with $|S| \ge 2$,
the marginal cost is reduced to $c_S := c -\epsilon_S$ 
when all firms in $S$ merge
where $c \ge \epsilon_S \ge 0$.
Note that synergy effects are heterogeneous depending on $S$.\footnote{
\citet{Lardon2019Core} considers a different model with synergy effects, 
where firms initially have heterogeneous marginal costs, 
and mergers reduce these costs to the lowest marginal cost within the coalition.
}
For instance, firms that are geographically close may benefit more from cost reductions due to lower transportation costs and easier coordination.
We assume that synergy effects are monotone; that is, $\epsilon_T \ge \epsilon_S$
for all $S, T$ such that $|S| \ge 2$ and $S \subseteq T$.
For each singleton set $\{i\} \subseteq N$, we denote $c_{ \{ i \} } := c$.
Then, the joint profit of a nonempty $S \subseteq N$ is
given by
$$
\pi_{S} (p)  = \sum_{i \in S} D(p)(p_i -c_S).
$$
for all $(p_1, \ldots, p_n) \in \mathbb{R}_{+}^n$.

We consider the $\alpha$-characteristic function as in \citet{Lardon2020Convexity}.\footnote{
The $\alpha$-characteristic function has the max--min form in~\eqref{eq:alphav}, whereas the $\beta$-characteristic function reverses the order of optimization and has the min--max form. \citet{Lardon2020Convexity} shows that the two characteristic functions coincide in this Bertrand setting.
\citet{Lardon2020Convexity} also considers the $\gamma$-characteristic function, under which the outsiders choose their prices as individual best responses to the coalition's action, and shows that the resulting game is convex. We focus on the $\alpha$-characteristic function because it makes the analysis tractable when merger synergies are introduced.

}
Specifically, the value of each coalition $S$, denoted by $\nu(S)$, is defined
as the maximal profit attained by $S$ when the outsiders (firms in $N \setminus S$)
act against $S$ in the worst possible way:
\begin{align}\label{eq:alphav}
\nu(S) = \max_{p_S \in \mathbb{R}_{+}^S} \min_{p_{N \setminus S} \in \mathbb{R}^{N \setminus S}_{+} } \pi_S(p_S, p_{N \setminus S}).
\end{align}
We assume that the demand intercept is sufficiently 
large so that $V > (1+r) c$.
Under this assumption, 
the result of \citet{Lardon2020Convexity} implies that an interior solution  to \eqref{eq:alphav} exists and is given by 
$$
p_i = \frac{1}{2} \left(c_S + \frac{V}{\delta_S} \right)
$$
for all $i \in S$ where $\delta_S = 1 + \frac{r (n- |S|)}{n} $.\footnote{The outsiders set their prices to zero, $p_{N \setminus S} = 0$.} Letting
$
p_S \equiv \frac{1}{2} \left(c_S + \frac{V}{\delta_S} \right),
$
the $\alpha$-characteristic function is given by
$$
\nu(S) = 
\begin{cases}
|S|(V - \delta_S p_S) (p_S  - (c - \epsilon_S) ) 
& \text{ if } |S| \ge 2 \\
(V - \delta_S p_S) (p_S  - c ) 
& \text{ if } |S| =  1 \\
0  & \text{ if } S = \emptyset.
\end{cases}
$$
 Note that $(N,\nu)$ is not symmetric in general as $\epsilon _S$ depends on $S$.

Let $v$ be the above TU game without synergy effects; that is,
 $\epsilon_S = 0$ for all $S \subseteq N$. In this case, the value of each coalition $S$ is given by
 $
 v(S) = |S| (V - \delta_{S} p^*)  (p^* -c) 
 $
for all $S \subseteq N$
where $
p^* = \frac{1}{2} \left(c  + \frac{V}{\delta_S} \right).
$
 \citet{Lardon2020Convexity} shows that $v$ is strictly convex.
However, this result no longer holds when synergy effects are present.
Intuitively, there is a jump in the marginal contribution when a firm joins a singleton coalition, compared with the case without synergy effects.
On the other hand, for certain levels of synergy, convexity is violated while $3$-distant convexity is preserved.
The following example illustrates this fact.

\begin{example}\label{ex:oligoply}
Let $n = 4$, $V=5$, $c= 2$, $r= \frac{1}{2}$ and $\epsilon_{S} = \frac{1}{2}$
for all $S \subseteq N$ with $|S| \ge 2$.
Then, $\nu$ depends only on $|S|$.
Figure \ref{fig:bertrand_marginal} illustrates the marginal contribution,
where the numerical values are rounded to four decimal places.

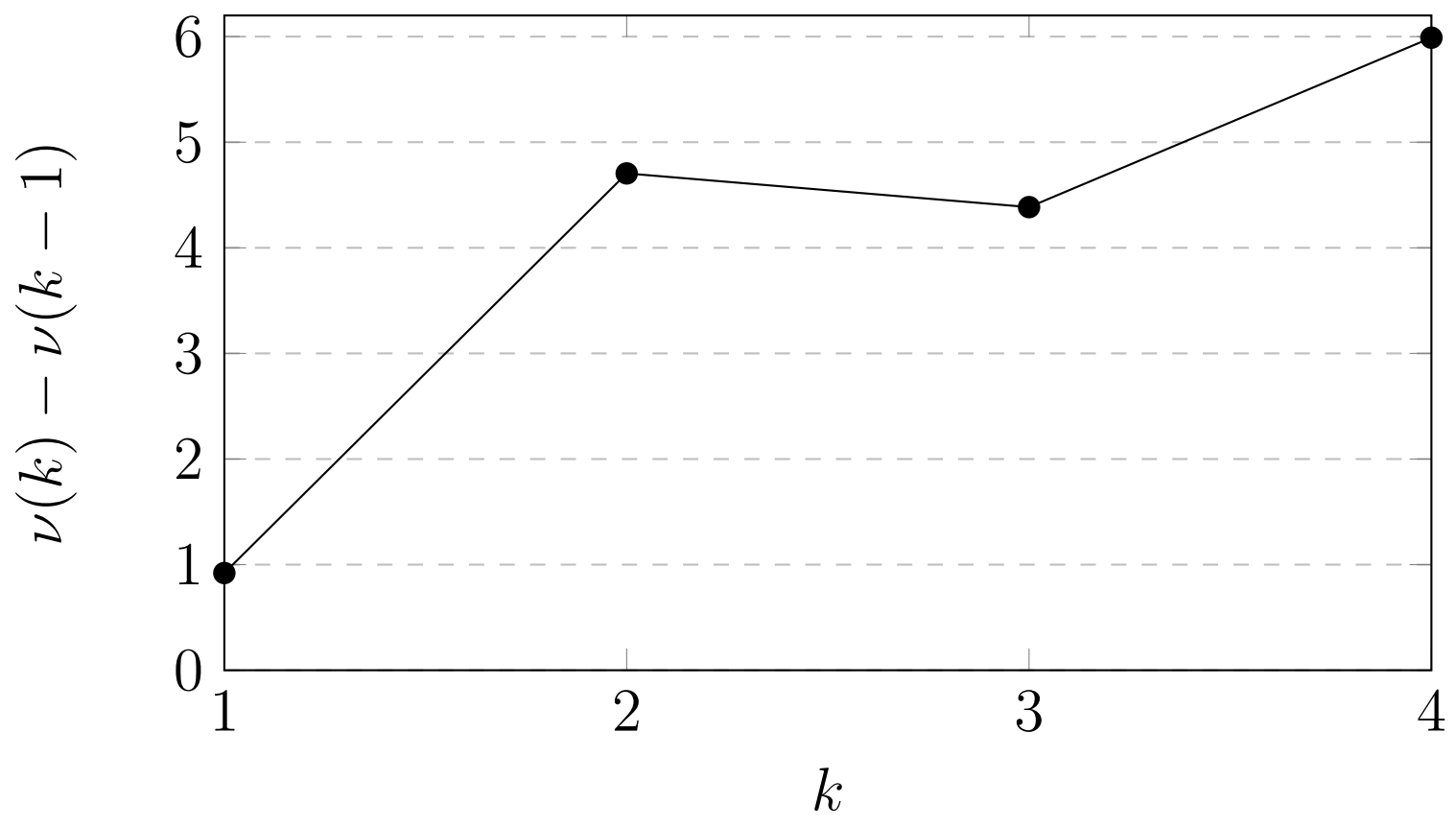
\begin{figure}[htbp]
\centering
\begin{tikzpicture}
\begin{axis}[
    width=0.65\textwidth,
    height=0.40\textwidth,
    xlabel={$k$},
    ylabel={$\nu(k)-\nu(k-1)$},
    xmin=1,
    xmax=4,
    ymin=0,
    ymax=6.2,
    xtick={1,2,3,4},
    ytick={0,1,2,3,4,5,6},
    ymajorgrids=true,
    xmajorgrids=false,
    grid style={dashed,gray!50},
    axis lines=box,
]
\addplot[
    black,
    thin,
    mark=*,
    mark options={fill=black},
]
coordinates {
    (1,0.9205)
    (2,4.7045)
    (3,4.3854)
    (4,5.9896)
};
\end{axis}
\end{tikzpicture}
\caption{Marginal contributions
$\nu(k)-\nu(k-1)$ for $k=1,\ldots,4$.}
\label{fig:bertrand_marginal}
\end{figure}

Thus, we have
$$
\nu(1) - \nu(0) < \nu(3) -\nu(2) < \nu(2) -\nu(1) < \nu(4) - \nu(3),
$$
which implies that $\nu$ is not convex while it is $3$-distant convex.

\end{example}

We provide a range of synergy effects under which $3$-distant convexity is satisfied.
As a benchmark, we first consider the case where synergy effects are constant across coalitions; 
that is, there exists $\epsilon > 0$ such that $\epsilon_S = \epsilon$ for all $S \subseteq N$ with $|S| \ge 2$.
In this case, we can characterize the threshold level of $\epsilon$ under which $3$-distant convexity (or convexity) is satisfied.
Let $M_2 = v(3) - v(2) - (v(2) -v(1)) > 0$ and $M_3 =v(4) - v(3) - (v(2) -v(1))$. By strict convexity of $v$, $M_3 > M_2 > 0$.
Let also $L_2 = 1 + \frac{r (n+1)}{n}$ and
$L_3 = 1 + \frac{r (n+3)}{n}$.
Define 
$\epsilon_k = \frac{ (c L_k -V) + \sqrt{ (cL_k -V)^2 + 4 L_k M_k}}{L_k}$ ($k =2, 3$). 

\begin{proposition}\label{prop:symmetric_oligopoly}
Suppose that the synergy effects are constant. Then,  
\begin{itemize}
\item[(a)] $0 < \epsilon_2 < \epsilon_3$ 
\item[(b)] $\nu$ is convex if and only if $\epsilon \le \epsilon_2$
\item[(c)] $\nu$ is 3-distant convex if and only if $\epsilon \le \epsilon_3$.
\end{itemize}
\end{proposition}
By $0 < \epsilon_2 < \epsilon_3$,
there always exists a range where convexity is violated while $3$-distant convexity is
satisfied.  In Example \ref{ex:oligoply},  we obtain 
$\epsilon_2 \approx 0.8014$ and $\epsilon_3 \approx 1.6826$, and thus there is a sufficiently large gap between the two values.


We next provide a sufficient condition for 3-distant convexity/convexity in general cases.
We further assume that synergy effects satisfy the following condition.
\begin{assumption}\label{weak-supermodular}
For all $i \in  N$ and  $S, T \subseteq N \setminus \{i\}$ with $S \subseteq T$ and $|S| \ge 2$, $$(\epsilon_{T \cup \{i\} } - \epsilon_T) \ge \frac{n-2}{n-1} (\epsilon_{S \cup \{i\} } - \epsilon_S).$$
\end{assumption}
Clearly, this condition is weaker than convexity of $\epsilon_S$. 
Then, we have the following result.
\begin{proposition}\label{prop:asymmetry_oligopoly}
Suppose that synergy effects satisfy Assumption 1.
\begin{itemize}
\item[(a)] If $\epsilon_{S} \le  \epsilon_2$ for all $S \subseteq N$ with $|S| =2$, then $\nu$ is convex.
\item[(b)]  If $\epsilon_{S} \le  \epsilon_3$ for all $S \subseteq N$ with $|S| =2$, then $\nu$ is $3$-distant convex.
\end{itemize}
\end{proposition}

We discuss the non-emptiness of the core.
In this model, the nonemptiness of the core is easily established by showing that the equal-split allocation lies in the core (see Appendix).
Specifically, the allocation $x$ with $x_i = \frac{\nu(N)}{n}$ for all $i \in N$ always belongs to the core.
However, this allocation treats all firms symmetrically and ignores heterogeneous synergy effects.
More precisely, the equal-split allocation is independent of the values $\epsilon_S$ ($S \subsetneq N$).
On the other hand, the $3$-distant convexity of the game enables us to construct a core allocation based on marginal contributions, and thus reflecting the entire structure of synergy effects via the extended greedy algorithm.
This class of games thus highlights the importance of analyzing
$3$-distant convexity beyond the question of core nonemptiness.

\subsection{Production games}
We consider a situation where players are involved in a project that requires two complementary tasks.  
 When a coalition is formed, the players are endogenously allocated between two groups to work on one of the tasks.  
For example, some workers mine iron ore and the rest mine coal to produce iron.  Another example is a workforce allocation in an R\&D division where some workers are assigned to an experimental team while the rest are assigned to a data analysis team. 

Each player $i$ is endowed with a labor resource $w_i>0$ (available time, skills, etc.) that can be used for either task.  We assume that the labor resource is indivisible, that is, each player can work on only one task.  
We say that players are \emph{similar} if $w_i+w_j>w_k$ for any distinct players $i,j,k$.  
We denote $w(S)=\sum _{i\in S}w_i$ for each $S\subseteq N$.  

The value of the project depends on how labor resources are allocated between two tasks and is determined by the production function $y=(x_1x_2)^r$, where $x_{\ell }$ is the total amount of the labor resource for task $\ell =1,2$ and $1/2<r<1$.  Since the function is homogeneous of degree $2r(>1)$, the project exhibits increasing returns to scale, where $r$ determines the degree of increasing returns to scale.
   
The worth of a coalition $S$ is determined by    
$v(S)=\max _{S'\subseteq S}\left(w(S')w(S\setminus S')\right)^{r}$ for each  $S \subseteq N$.  
We refer to $(N,v)$ as a \emph{production game}.  
For each $S \subseteq N$, we denote $f^1(S)\in \text{arg}\max _{S'\subseteq S}\left(w(S')w(S\setminus S')\right)^{r}$ and $f^2(S)=S\setminus f^1(S)$. We assume that $w(f^1(S))\le w(f^2(S))$ for each $S \subseteq N$ without any loss of generality.  
Note that $f^1(S),f^2(S)\in \text{arg}\max _{S'\subseteq S}w(S')w(S\setminus S')$ for each $S \subseteq N$.  This game is a variant of a production game with increasing returns to scale by \citet{IU1993IJGT}, who assume that the amount of resources available to each player for each task is fixed.  


The following example illustrates that  $v$ may not be convex  when $r < 1$.
\begin{example}\label{Eg:production}
Let $N  = \{1, \ldots, 8\}$, $r = 0.8$, and
$w_i = 1$ for all $i \in N$.
Then, $v$ depends only on $|S|$.
In this case, $v(S)$ is attained by distributing players equally to each task.
Therefore,
$$
v(k) = 
\begin{cases}
(\frac{k}{2} \cdot \frac{k}{2} )^{0.8} & \text{if $k$ is even} \\
\left(\frac{(k - 1)}{2} \cdot (k -\frac{(k-1)}{2} \right)^{0.8} & \text{if $k$ is odd}. 
\end{cases}
$$
Figure \ref{figure:production} illustrates the marginal contribution,
where the numerical values are rounded to four decimal places.
Clearly, $v$ is not convex since, for example, $v(2) - v(1) > v(3) - v(2)$.
On the other hand, $v$ is $3$-distant convex since
 $v(k + 1) - v(k) <  v(k +3 )- v(k+2)$ for all $k =0, \ldots, 5$.
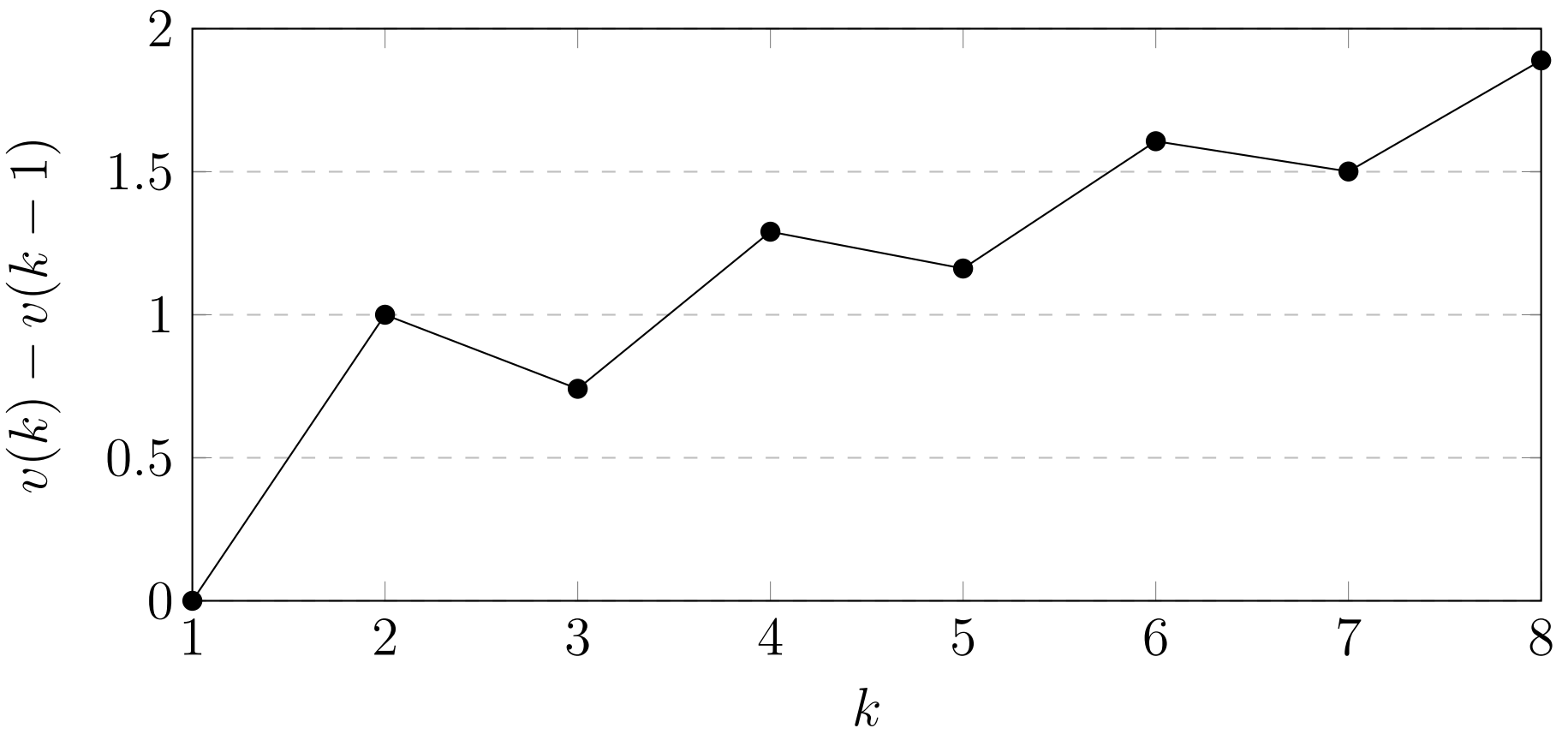
\begin{figure}[htbp]
\centering
\begin{tikzpicture}
\begin{axis}[
    width=12cm,
    height=6cm,
    xlabel={$k$},
    ylabel={$v(k)-v(k-1)$},
    xmin=1, xmax=8,
    ymin=0, ymax=2,
    xtick={1,2,3,4,5,6,7,8},
    ymajorgrids=true,
    grid style=dashed,
]
\addplot[mark=*] coordinates {
    (1,0.0000)  
    (2,1.0000)
    (3,0.7411)
    (4,1.2903)
    (5,1.1616)
    (6,1.6066)
    (7,1.5008)
    (8,1.8892)
};
\end{axis}
\end{tikzpicture}

\caption{Marginal contributions $v(k)-v(k-1)$ for $k = 1, \ldots, 8$.}\label{figure:production}
\end{figure}
\end{example}

In general, the following proposition holds.

\begin{proposition}\label{prop:production}
If $r$ is sufficiently close to $1$  and the players are similar, then the  production game is 3-distant convex.
\end{proposition}
 Therefore, we can check the nonemptiness of the core by \eqref{ineq:ns_core} in Theorem \ref{thm:ns_core} and find a core  allocation when one exists by using the algorithm provided in Theorem  \ref{thm:algorithm}.

\begin{remark}
    {\rm Proposition \ref{prop:production} remains valid for the case where $r\ge 1$, although the production game turns out to be convex in this case.  The assumption $r<1$ is imposed to capture the decreasing marginal productivity of each task when the amount of the labor input for the other task} is fixed. 
\end{remark}

\section{Conclusion  }\label{sec:extension}

This paper has investigated the core of $3$-distant convex games.
We have provided a necessary and sufficient condition for core
nonemptiness and developed a strongly polynomial-time combinatorial
algorithm for finding a core allocation whenever the core is
nonempty. Our applications demonstrate that $3$-distant convexity
can capture economically relevant settings in which convexity fails
only locally.

Several directions remain for future research. First, it would be
natural to extend our analysis to $k$-distant convex games for
$k\geq 4$. As $k$ increases, the convexity condition is imposed on
fewer pairs of coalitions, allowing a wider range of local departures
from convexity. Such extensions may therefore broaden the scope of
economic applications covered by the framework. At the same time,
the structural arguments developed for the case $k=3$ may no longer
apply directly. It would be interesting to identify conditions for
core nonemptiness and to develop efficient methods for constructing
core allocations in these more general classes.

Second, while this paper focuses on the core, other solution concepts
for $3$-distant convex games also merit investigation. Possible
directions include studying the Shapley value, the nucleolus, and the
least core. In particular, it would be useful to understand how local
violations of convexity affect the properties and computational
tractability of these solutions, and which of the familiar results
for convex games continue to hold under $3$-distant convexity.






\section*{AI disclosure}
We used OpenAI's GPT models and Anthropic's Claude models to assist in developing some of the proofs and improving the presentation of the manuscript.
In particular, the proofs of Propositions \ref{prop:symmetric_oligopoly} and \ref{prop:asymmetry_oligopoly} were obtained with the assistance of ChatGPT.
The authors take full responsibility for the content of this paper, including all mathematical claims and proofs.

\section*{Appendix A.  Omitted Proofs in Section  \ref{sec:main}}

\subsection*{Proof of Lemma \ref{claim:core0} }

We first provide a basic property of $3$-distant convex games.
\begin{lemma}\label{lemma:3convex1}
Let $(N, v)$ be a 3-distant convex game.
For any $S \subseteq N$ and $i \in S$, if
$v(S ) - v(S \setminus \{i\}) > v(N)  - v(N \setminus \{i\})$,
then $S = N \setminus \{j\}$ for some $j \in N \setminus \{i\}$.
\end{lemma}
\begin{proof}
 Let $S \subseteq N$ and $i \in S$.
 Suppose that $v(S ) - v(S \setminus \{i\}) > v(N)  - v(N \setminus \{i\})$.
If $|S| < |N| -1$,
then $|N \setminus \{i\}| - |S \setminus \{i\}| = |N| -|S| \ge 2$,
contradicting Proposition \ref{prop:marginal}.
Thus, $|S| = |N| -1$ by $S \neq N$.
By $i \in S$, $S = N \setminus \{j\}$
for some $j \in N \setminus \{i\}$.

\end{proof}

We now show Lemma \ref{claim:core0}.
Note that $x(N \setminus \{ \hat{i}\}) = v(N \setminus \{\hat{i} \})$
since $(x_{i})_{i \in N \setminus \{ \hat{i} \}}$ is a core allocation for $(N \setminus \{ \hat{i} \}, v_{-\hat{i}})$.
By $x_{ \hat{i} } = v(N) - v(N \setminus \{\hat{i} \})$,
we have $x(N) = v(N)$.
Suppose that $x$ is not in the core.
Then, there exists a nonempty $S \subseteq N$ with $v(S) > x(S)$.
We must have $\hat{i} \in S$ since $(x_{i})_{i \in N \setminus \{ \hat{i} \}}$
is a core  allocation for $(N \setminus \{ \hat{i} \}, v_{-\hat{i}})$.
We also have $x(S \setminus \{ \hat{i} \}) \ge v(S \setminus \{ \hat{i} \})$.
Thus, $v(S) - v(S \setminus \{ \hat{i} \}) > x_{ \hat{i} }$ by $v(S) > x(S)$.
By Lemma \ref{lemma:3convex1}, $S = N \setminus \{j\}$ for some $j \in N \setminus \{ \hat{i} \}$.

\subsection*{Proof of Lemma \ref{claim:core1} }

(a) As $x$ is not in the core,
there exists a nonempty $S \subseteq N$ with $v(S) > x(S)$ such that 
 $S = N \setminus \{j\}$ for some $j \in N \setminus \{ \hat{i} \}$ by Lemma \ref{claim:core0}. Thus $v(N \setminus \{j\}) > x(N \setminus \{j\})$.
By $v(N) = x(N)$, we have $x_j > v(N) -  v(N \setminus \{j\})$.
Therefore, $j \in N^{+}$.
If $N^{-} = \emptyset$, we have
$
x(N) > x^{*}(N) \ge v(N)
$
by (\ref{ineq:ns_core}), a contradiction.

(b) Since $(N, v)$ satisfies (\ref{ineq:ns_core}),
we have  
$$ 
x^{*}(N^{+})  + x^{*}(N^{=}) + x^{*}(N^{-}) =  x^{*}(N) \ge  v(N) = x(N^{+}) + x(N^{=}) + x(N^{-})
$$
Equivalently, $ x(N^+) - x^{*}(N^{+} ) \le x^*(N^{-}) - x(N^{-}) $.

\subsection*{Proof of Lemma \ref{claim1:case1_core} }

Note that
$x'(N)=x^*(N^{+}) + x(N^{=}) + x'(N^{-}) = x(N)$
since  $x(N^{+})  - x^{*}(N^{+}) =  x'(N^{-}) - x(N^{-})$ by the definition of $x'$.
Moreover, we have already shown $x(N) = v(N)$ in the proof of Lemma \ref{claim:core1}.
Therefore, $x'(N) = v(N)$.

Suppose that $x'$ is not in the core.
Then, there exists a nonempty $S \subseteq N$
such that $v(S) > x'(S)$.  
Take a minimal $S$ that satisfies $v(S) > x'(S)$.
By the minimality of $S$,  $v(S\setminus \{ i\} )\le x'(S\setminus \{ i\} )$ for any $i\in S$.  Thus, we have that for any $i \in S$,
\begin{align}
v(S) - v( S \setminus \{ i \} ) > x'_{i}.\label{claim1:eq1}
\end{align}

Suppose that $S \subseteq N^{-}$.
Then, $S \subseteq N \setminus \{ \hat{i} \}$ by $\hat{i} \in N^{=}$.
By the definition of $x'$, 
$ x'(S) \ge x(S)$.
Moreover, $x(S) \ge v(S)$ since $x$ is a core allocation for $(N \setminus \{ \hat{i} \}, v_{-\hat{i}})$, contradicting $v(S) > x'(S)$.

Suppose that $S \nsubseteq N^{-}$.
Pick any $ i \in S  \setminus N^{-}$.
Then, $i \in N^+ \cup N^{=}$ and thus $x'_i = v(N) - v(N \setminus \{i\} )$ by definition.
By (\ref{claim1:eq1}) and Lemma \ref{lemma:3convex1},
we have $S = N \setminus \{\hat{j} \}$ for some $\hat{j} \in N \setminus \{i\}$.
Thus, $v(N \setminus \{\hat{j} \}) > x'(N \setminus \{\hat{j} \}) $,
which implies $\hat{j} \in N^{-}$ since $v(N)=x'(N)$ and $x'_i=v(N)-v(N\setminus \{ i\} )$ for all $i\in N^+\cup N^=$.
Therefore,
\begin{align*}
v(N \setminus \{\hat{j} \}) &>  x'(N) - x'_{ \hat{j }} \\
&= x(N) - x'_{ \hat{j }} \\
& = x_{\hat{i}} + x(N \setminus \{\hat{i}\})  - x'_{ \hat{j }} \\
& = x_{\hat{i}} + v(N \setminus \{\hat{i}\})   - x'_{ \hat{j }} \\
& = v(N) -  x'_{\hat j},
\end{align*}
where the second line follows from that $x(N) = x'(N) = v(N)$, the fourth line follows
from  the fact that $x$ is a core  allocation for $(N \setminus \{\hat{i}\}, v_{-\hat{i}})$, and the fifth line follows from $x_{\hat i}=v(N)-v(N\setminus \{ \hat i\} )$.
Thus, $x'_{ \hat{j} } > x^{*}_{ \hat{j} }$, contradicting the construction of $x'$ by $\hat j\in N^-$.




\subsection*{Proof of Lemma \ref{claim:imputation} }

 We begin with a basic lemma.
\begin{lemma}\label{lemma:3convex0}
Let $(N, v)$ be a $0$-nomalized and 3-distant convex game.
For any $S, T \subseteq N$ with $S \subseteq T$ and $|S| \ge 2$, we have $v(S) \le v(T)$.
\end{lemma}
\begin{proof}
 It suffices to show the case where $S\subsetneq T$. Let $\{i_1, i_2, \ldots, i_k\} =T \setminus S$.
As $v$ is $0$-normalized and 3-distant convex,
$v(S) = v(S) + v(\{i_1\}) \le v(S \cup \{i_1\}).$
By repeating this argument,
we have $v(S) \le v(T)$.
\end{proof} 
Moreover, the following two lemmas are useful for the proof. Recall that $x^*_i = v(N) - v(N \setminus \{i\})$ for all $i \in N$.
Note that  $(N,v)$ satisfies (\ref{ineq:ns_core}) if and only if $ x^*(N) \ge v(N)$.  
 Note also that $x^*_i\ge 0$ for all $i\in N$ since $3$-distant convexity of $v$ and $n\ge 3$ imply $v(N)\ge v(N\setminus \{ i\} )+v(\{ i\} )=v(N\setminus \{ i\} )$ for all $i\in N$. 
\begin{lemma}\label{lemma:marginal}
Let $(N, v)$ be a 3-distant convex game.
Let $i_1,\ldots,i_n$ be any ordering of players and $m$ be a marginal worth vector corresponding to $i_1,\ldots,i_n$.
Then, $x^*_{i} \ge m_{i}$ for all $i  \in N \setminus \{i_{n-1} \}$.
\end{lemma}
\begin{proof}
By definition, $m_{i_n} = x^{*}_{i_{n} }$. 
Consider any $i_{\ell} $ with $1 \le \ell \le n-2$.
By $\ell \le n-2$,
we have $|\{i_1,\ldots, i_{\ell-1} \}|
+ 2
\le |N| -1 $.
By Proposition \ref{prop:marginal},
we have
$x^*_{i_{\ell}} =
v(N) - v(N \setminus \{i_{\ell} \})
\ge v(\{i_1,\ldots, i_{\ell-1}, i_{\ell} \} ) - v(\{i_1,\ldots, 
i_{\ell-1} \}) = m_{i_{\ell}}.
$
\end{proof}

\begin{lemma}\label{lemma:3convex2}
Let $(N, v)$ be a 3-distant convex game.
Suppose that $x^*(N) = v(N)$.
Then, $x^*$ is in the core.
\end{lemma}
\begin{proof}
Suppose that $x^*$ is not in the core.
Then, $v(S) > x^*(S)$ for some nonempty $S \subseteq N$. 
Take a minimal $S$ with  $v(S) > x^*(S)$.  
Let $i \in S$.
 By the minimality of $S$, $v(S\setminus \{ i\} )\le x^*(S\setminus \{ i\} )$.  
Then, $v( S ) -  v(S \setminus \{i\}) > x^*_i = v(N) -v(N \setminus \{i\}) $.  
By Lemma \ref{lemma:3convex1}, $S = N \setminus \{j \}$ for some $j \in N \setminus \{i\}$.
Thus, $v(N \setminus \{j\})   > x^*( N \setminus \{j\})$.
By adding $ v(N) - v(N \setminus \{j\}) = x^{*}_j$,
we have $v(N) > x^{*}(N)$, a contradiction.
\end{proof}

 We turn to the proof of Lemma \ref{claim:imputation}.
Suppose that $v(N \setminus \{\hat{i}\}) \ge 0$ and 
$(N \setminus \{ \hat{i} \}, v_{-\hat{i}})$ does not satisfy (\ref{ineq:ns_core}).
Let $y^{*}_i = v(N \setminus \{ \hat{i} \}) - v(N \setminus \{ \hat{i}, i \})$
for all $i \in N \setminus \{ \hat{i} \}$.
Recall that $|N| = n$ and  
\begin{align}
\epsilon = \frac{ \sum_{i \in N \setminus \{\hat{i}\} } v(N \setminus \{\hat{i}, i\}) }{n-2} -  v(N \setminus \{ \hat{i} \} )  > 0, \label{eq:case1}
\end{align}
equivalently,
\begin{align}
\epsilon = \frac{1}{n-2} (v (N \setminus \{ \hat{i} \}) - \sum_{i \in N \setminus \{ \hat{i} \} } y^{*}_i ) > 0. \label{eq:case2}
\end{align}

Note that $n \ge 4$ holds in this case
since 
 (\ref{eq:case1}) implies $v(N \setminus \{\hat{i}\}) < 0$ when $n = 3$ as $v$ is $0$-normalized.   Therefore, we assume $n\ge 4$ in the remainder of the proof. 
Define $(N \setminus \{ \hat{i} \}, \nu)$ by
$$
\nu(S) = 
\begin{cases}
v(N \setminus \{ \hat{i} \}) + \epsilon  & \text{ if } S =  N \setminus \{ \hat{i} \}, \\
v(S)   & \text{ if } S \neq  N \setminus \{ \hat{i} \}.\\
\end{cases}
$$

Then, it is straightforward to see that $(N \setminus \{ \hat{i} \}, \nu)$ 
is $0$-normalized, $\nu(N \setminus \{\hat{i}\}) \ge 0$, and $3$-distant convex.
By \eqref{eq:algorithm2},
$y_i = \nu(N \setminus \{\hat{i}\}) - \nu(N \setminus \{\hat{i}, i\})$
for each $i \in N \setminus \{ \hat{i} \}$.
By the definition of $\epsilon$, we have
$$
\nu(N \setminus \{ \hat{i} \}) = \sum_{i \in N \setminus \{ \hat{i} \} } y_i.
$$
By Lemma \ref{lemma:3convex2}, $(y_i)_{i \in N \setminus \{ \hat{i} \} } $ is a core allocation
for  $(N \setminus \{ \hat{i} \}, \nu)$.

To complete the proof, we show that  $y=(y_i)_{i\in N}$ is a core allocation
for $(N, v)$.  Recall that $y_{ \hat{i} } = v(N) - v(N \setminus \{ \hat{i} \}) - \epsilon$.
By definition, $y(N) = v(N) $.  

 We first show that $y_i \ge v(\{ i\} )=0$ for all $i\in N$.  The inequality holds for all $i \in N \setminus \{ \hat{i} \}$ 
 since $(y_i)_{i \in N \setminus \{ \hat{i} \} } $ is a core allocation
for  $(N \setminus \{ \hat{i} \}, \nu)$.
 
Suppose that $y_{ \hat{i} } < 0$.
Let  $i_1, \ldots, i_{n-1}, i_n$ be an  ordering of $N$ such that $i_{n} = \hat{i}$
and  $m$ be the marginal worth vector for $(N, v)$ corresponding to this order.
Note that $m_{i_{n-1} } = y^{*}_{ i_{n-1} }$  and $\sum _{\ell =1}^{n-1}m_{i_{\ell }}=v(N\setminus \{ \hat i\} )$ by definition.
As $(N \setminus \{ \hat{i} \}, v_{-\hat{i}})$ is 3-distant convex,
  $m_{i } \le y^{*}_{ i }$ for any $i \in N \setminus \{ \hat{i}, i_{n-2} \}$ from Lemma \ref{lemma:marginal}.
By $y_{ \hat{i} } < 0$ and (\ref{eq:case2}), 
\begin{align*}
0  & >  (n-2)y_{\hat i} \\
& = (n-2) m_{ i_{n} } - \left(v(N\setminus \{ \hat i\} )-\sum _{i\in N\setminus \{ \hat i\} }y^*_i\right)  \\ 
& = (n-2) m_{ i_{n} } -  \sum^{n-1}_{ \ell = 1} ( m_{i_{\ell}} - y^{*}_{ i_{\ell} } ) \\
& \ge m_{ i_{n} }  - (m_{ i_{n-2} } - y^{*}_{ i_{n-2} } )  \\
&  = v(N) - v(N \setminus \{i_n\}) -  m_{i_{n-2}}  +  v(N \setminus \{i_n \}) -  v(N \setminus \{i_n, i_{n-2} \}) \\
& = v(N) -  v(N \setminus \{i_n, i_{n-2} \})  - m_{i_{n-2}} \\
& \ge v(N) - v(N \setminus \{i_{n-2} \})  - m_{i_{n-2}},
\end{align*}
where the fourth line follows from the fact that $n \ge 4$, $m_{i_n} = v(N)- v(N \setminus \{ \hat{i} \}) \ge 0$ and $m_{i } \le y^{*}_{ i }$ for all $i \in N \setminus \{ \hat{i}, i_{n-2} \}$,
and the final line follows from Lemma \ref{lemma:3convex0}
and $n \ge 4$.
Therefore, we have
$$
v(\{i_1, \ldots, i_{n-3}, i_{n-2} \}) - v(\{i_1, \ldots, i_{n-3} \}) > v(N)  - v(N \setminus \{i_{n-2} \}),
$$
contradicting that $v$ is $3$-distant convex by Proposition \ref{prop:marginal}.
 Hence, $y_{ \hat{i} } \ge 0$. 

We now show that $y(S)\ge v(S)$ for any coalition $S$. 
Suppose that 
there exists some $S \subsetneq N$ such that $v(S) > y(S)$.  
Suppose that $\hat{i} \notin S$.
By the definition of $\nu$, $\nu(S)  \ge v(S) > y(S)$, contradicting
that $(y_i)_{i \in N \setminus \{ \hat{i} \} } $ is a core allocation
for  $(N \setminus \{ \hat{i} \}, \nu)$.

Therefore, $\hat{i} \in S $.
Moreover, $|S| \ge 2$ since $y_i \ge 0$ for all $i \in N$.
We denote $S = \{i_1, \ldots, i_{m} \}$ where  $i_m = \hat{i}$.

We first assume that $m = 2$.
By $v(S)  > y(S)$, we have
\begin{align*}
v( \{ i_1, i_2\} ) & > y_{i_1} + y_{i_2} \\
& = v(N \setminus \{i_2\} ) + \epsilon - v(N \setminus \{ i_2, i_1\} )  + v(N) - v(N \setminus \{i_2\} ) - \epsilon \\
& = - v(N \setminus \{i_2, i_1\} )  + v(N).
\end{align*}
Thus, $v(N \setminus \{i_1, i_2\}) + v(\{i_1, i_2\}) > v(N) $, contradicting  that $v$ is $3$-distant convex by $N \setminus \{i_1, i_2 \} \neq \emptyset$.

We next assume that $m \ge 3$.
By $v(S)  > y(S)$ and $\epsilon >0$, we have
\begin{align*}
v( S) & > y(S \setminus \{i_m\}) + y_{i_m}  \\
& = (m-1) \epsilon + (m - 1) v(N \setminus \{i_m\}) - \sum^{m-1}_{\ell =1} v(N \setminus \{i_m, i_{\ell}  \} ) + v(N) - v(N \setminus \{i_m \} )  - \epsilon \\
& >  (m -2) v(N \setminus \{i_m\}) - \sum^{m-1}_{\ell =1} v(N \setminus \{i_m, i_{\ell}  \} ) +v(N).
\end{align*}
Therefore, 
\begin{align}
v(N)  + (m -2) v(N \setminus \{i_m\}) < 
v( S) + \sum^{m-1}_{\ell =1} v(N \setminus \{i_m, i_{\ell}  \} ).\label{eq:case2_1}
\end{align}

 We denote $T = N \setminus \{i_m, i_1\}$.
Then, $i_m, i_1 \in S \setminus T$.
Moreover, $|T \setminus S| \ge 1$ by $\emptyset \neq N \setminus S = T \setminus S$.
As $v$ is $3$-distant convex, 
$$v(S) + v(N \setminus \{i_m, i_1\}) \le v(S \cup T) + v(S\cap T) = v(N) + v(S \setminus \{i_m, i_1\} ).
$$
Therefore,
\begin{align}
v( S) + \sum^{m-1}_{\ell =1} v(N \setminus \{i_m, i_{\ell}  \} ) \le
v(N) + v(S \setminus \{i_m, i_1 \}) +  \sum^{m-1}_{\ell =2} v(N \setminus \{i_m, i_{\ell}  \} ) \label{eq:case2_2}
\end{align}

 We next show that  for each $\ell = 1, \ldots, m-2$,
\begin{align}
v(S \setminus \{i_m, i_1, \ldots, i_{\ell} \}) +  v( N \setminus \{i_m, i_{\ell +1 }\} )
\le v(N \setminus \{i_m \}) +  v( S \setminus \{i_m, i_{1}, \ldots, i_{\ell}, i_{\ell +1} \} ).\label{eq:case2_3}
\end{align}
Let $\ell = 1, \ldots, m-2$.
By $S = \{i_1, \ldots, i_{m} \}$ and $|S| <  n$,
$|S \setminus \{i_m, i_1, \ldots, i_{\ell+1} \}| + 2
= m - (\ell +2) + 2 < n - \ell \le n-1$.
Thus, $|S \setminus \{i_m, i_1, \ldots, i_{\ell+1} \}| + 2 \le n-2 = |N \setminus \{i_{m}, i_{\ell +1} \}|.
$
By Proposition \ref{prop:marginal},
$$
v(N \setminus \{i_m\}) - v(N \setminus \{i_m, i_{\ell + 1} \}) \ge
v(S \setminus \{i_m, i_1, \ldots, i_{\ell} \}) -
v( S \setminus \{i_m, i_{1}, \ldots, i_{\ell}, i_{\ell +1} \} ),
$$
which is equivalent to (\ref{eq:case2_3}).


By (\ref{eq:case2_2}) and (\ref{eq:case2_3}) for $\ell =1$,
\begin{align*}
v( S) + \sum^{m-1}_{\ell =1} v(N \setminus \{i_m, i_{\ell}  \} ) & \le
 v(N) + v(S \setminus \{i_m, i_1 \}) +  \sum^{m-1}_{\ell =2} v(N \setminus \{i_m, i_{\ell}  \} ) \\
&\le  v(N) + v(N \setminus \{i_m\}) + v(S \setminus \{i_m, i_1, i_2 \})  + 
\sum^{m-1}_{\ell =3} v(N \setminus \{i_m, i_{\ell}  \} ).
\end{align*}
By repeating this argument, we have that for any $\ell = 2, \ldots, m-2$,
\begin{align*}
v( S) + \sum^{m-1}_{\ell =1} v(N \setminus \{i_m, i_{\ell}  \} ) & \le v(N) + \ell  v(N \setminus \{i_m\}) + v(S \setminus \{i_m, i_1, \ldots, i_{\ell+1} \})  + 
\sum^{m-1}_{\ell' = \ell +2} v(N \setminus \{i_m, i_{\ell'}  \} ),
\end{align*}
where $\sum^{m-1}_{\ell' = m} v(N \setminus \{i_m, i_{\ell'}  \} ) \equiv 0$.
In particular, when $\ell = m-2$,
\begin{align*}
v( S) + \sum^{m-1}_{\ell =1} v(N \setminus \{i_m, i_{\ell}  \} ) & \le v(N) + (m - 2) v(N \setminus \{i_m\}) + v(S \setminus \{i_m, i_1, \ldots, i_{m-1} \}) \\
 & = v(N) + (m-2) v(N \setminus \{i_m\}),
\end{align*}
 contradicting (\ref{eq:case2_1}).

\subsection*{Proof of Lemma \ref{claim:imputation_minus}}


Clearly,  $z(N)=v(N)$ and $z_i\ge 0=v(\{ i\} )$ for all $i\in N$
by $v(N) \ge 0$.
Suppose that there exists $S \subsetneq N$ 
with  $v(S) > z(S)$.
As $v$ is $0$-normalized and  $z(S) \ge 0$,
$|S| \ge 2.$
When $\hat{i} \in S$,  $v(S) > v(N)$, contradicting Lemma \ref{lemma:3convex0}.
When $\hat{i} \notin S$,  $v(S) >  z(S) =  0 > v(N \setminus \{\hat{i} \} ) $, contradicting Lemma \ref{lemma:3convex0}.
Thus,  $z$ is a core allocation.

\section*{Appendix B. Omitted proofs in Section \ref{sec:applications}}

\subsection*{Proof of Propositions 
\ref{prop:symmetric_oligopoly} and \ref{prop:asymmetry_oligopoly} }




%
We assume that the synergy effects are non-decreasing ($\epsilon_{T} \ge \epsilon_{S}$ 
for all $S, T \subseteq N$ with $|S|\ge 2$ and $S\subseteq T$) and satisfy Assumption 1 throughout the proof.
 By direct calculation, we obtain the following expression for  $\nu(S)$ for all $S \subseteq N$ with $|S| \ge 2$:
\begin{align}
\nu(S) = v(|S|) - \frac{1}{4} ( 2 \epsilon_S c - \epsilon_{S}^2 ) |S| \delta_S + \frac{\epsilon_S V |S| }{2}. \label{eq:oligopoly0}
\end{align}
Moreover, for all $i \in N$ and $S \subseteq N \setminus \{i\}$ with $|S| \ge 2$, we have
\begin{align}
\nu(S \cup \{i\}) - \nu(S) = v(|S \cup \{i\}|) - v(|S|) + F_i(S), \label{eq:oligopoly1}
\end{align}
where
$F_i(S)$ can be written as
\[
F_i(S) = G_i(S) + H_i(S),
\]
with $G_i(S)$ and $H_i(S)$ given by
$$
G_i(S) = \frac{1}{4} |S| \left(\epsilon_{S \cup \{i\}} - \epsilon_S \right) \left(2V - \delta_S \left(2c -  (\epsilon_{S \cup \{i\}} + \epsilon_S) \right) \right) 
$$
and
$$
H_i(S) = -\frac{1}{4} (2 c \epsilon_{S \cup  \{i\}} - \epsilon^2_{S \cup \{i\}} )  \left( 1 + r \frac{ ( n - 2|S|  - 1)}{n} \right) +\frac{\epsilon_{S \cup \{i\}} V}{2}.
$$

We first introduce a technical lemma that is useful for subsequent analysis.
Define $f : \mathbb{R}^2 \to \mathbb{R}$ by
\[
f(x, y) = -\frac{1}{4} (2cx - x^2) y + \frac{V}{2} x.
\]
 Note that $H_i(S)=f(x,y)$ when $x=\epsilon _{S\cup \{ i\} }$ and $y=1+\frac{r(n-2|S|-1)}{n}$.
\begin{lemma}\label{lemma:oligopoly1}
\begin{itemize}
\item[(a)] $\displaystyle{\frac{\partial}{\partial x} f (x, y) \ge 0}$ for all $x \in [0, c]$ and  $y \le 1+r $.
\item[(b)]  $\displaystyle{ \frac{\partial}{\partial y} f (x, y) \le 0}$ for all $x \in [0, 2c]$ and $y \in \mathbb{R}$.
\end{itemize}
\end{lemma}
\begin{proof}
For part (a), we have
$$
    \frac{\partial}{\partial x} f (x, y)   = \frac{1}{2} \left( V - (c- x) y \right) 
     \ge \frac{1}{2} (V -(1+r) c ) 
      > 0,
$$
where the first inequality follows from $c - x \ge 0$ and $ y  \le  1+r$  and the strict inequality follows from the assumption that $V > (1+r)c$.
For part (b), we have
$
  \displaystyle{\frac{\partial}{\partial y} f (x, y)  } = - \frac{1}{4} x (2c -x) 
      \le 0
$
by $x \in [0, 2c]$.
\end{proof}

We next show that $G_i$, $H_i$, and $F_i$ are nonnegative and non-decreasing. 
Moreover, the marginal contribution of $\nu$ is non-decreasing except for $|S|=1$.
\begin{lemma}\label{lemma:oligopoly2}

\begin{itemize}
 \item[(a)] $G_i(T) \ge G_i(S) \ge 0$, $H_i(T) \ge H_i(S) \ge 0$, and $F_i(T) \ge F_i(S) \ge 0$ for all $ i \in N$ and $S, T \subseteq N \setminus \{i\}$ such that
 $|S| \ge 2$ and $S \subseteq T$, 
 \item[(b)] $\nu(T \cup \{i\}) - \nu(T) \ge \nu(S \cup \{i\}) - \nu(S)$
 for all $ i \in N$ and $S, T \subseteq N \setminus \{i\}$ such that
 $|S| \neq 1$ and $S \subseteq T$.
 \end{itemize}
\end{lemma}
\begin{proof}

(a) 
Fix any $i \in N$ and $S,  T \subseteq N \setminus \{i\}$  such that  $|S| \ge 2$ and $S \subseteq T$.
We first show that $G_i(S) \ge 0$.
Note that 
\begin{align}
2V -  \delta_S(2c - (\epsilon_{S \cup \{i\}}  + \epsilon_S)) &\ge 2V - 2c\delta_S  \ge 2 (V - (1+r)  c) > 0,  \label{eq:oligopoly2}
\end{align}
where the first inequality follows from $\delta_S \ge 0$ and $\epsilon_{S \cup \{i\}} +  \epsilon_S \ge 0$, the second inequality follows from $\delta _S=1+\frac{r(n-|S|)}{n}\le 1+r$,  and
the  strict inequality follows from the assumption that $V > (1+r)c $.
This implies $G_i(S) \ge 0$ as  the synergy effects are non-decreasing.

We next show that $H_i(S) \ge  0$.
Note that $ 1 + r \frac{( n - 2|S|  - 1)}{n} \le (1+r)$ and 
 $2c-\epsilon _{S\cup \{ i\} }\ge c$, where the latter follows from the assumption that $\epsilon _{S\cup \{ i\} }\le c$.
This implies
\begin{align*}
H_i(S) &= -\frac{1}{4} \epsilon_{S \cup  \{i\}} (2 c  - \epsilon_{S \cup \{i\}} )  \left( 1 + r\frac{ ( n - 2|S|  - 1)}{n} \right) +\frac{\epsilon_{S \cup \{i\}} V}{2} \\
& \ge -\frac{1}{2} \epsilon_{S \cup  \{i\}}   c (1+r) +\frac{\epsilon_{S \cup \{i\}} V}{2}  = \frac{\epsilon_{S \cup  \{i\}}}{2} (V - (1+r)c) \ge 0,
\end{align*}
where the last inequality follows from the assumption that $V > (1+r)c $.
Therefore, $F_i(S) = G_i(S) + H_i(S) \ge 0$.

 We next show that $G_i(T) \ge G_i(S)$.
Note that
\begin{align}
2V -  \delta_T\left(2c - (\epsilon_{T \cup \{i\} }  + \epsilon_T) \right) 
\ge 2V -  \delta_S \left(2c - (\epsilon_{S  \cup \{i\} }  + \epsilon_S) \right) > 0,
\end{align}
where the first inequality follows from $\delta_S \ge \delta_T \ge 0$ and 
$2c - (\epsilon_{S \cup \{i\}}+ \epsilon_{S}) \ge 2c - (\epsilon_{T \cup \{i\}}+ \epsilon_{T})  \ge 0 $, and the strict inequality follows from (\ref{eq:oligopoly2}).
Therefore,  when $\epsilon_{S \cup \{i\}} - \epsilon_S > 0$,  

\begin{align*}
\frac{G_i(T)}{ G_i(S)} & = \frac{|T|}{|S|} \frac{(\epsilon_{T \cup \{i\}} - \epsilon_T)}{(\epsilon_{S \cup \{i\}} - \epsilon_S) } \frac{\left(2V -  \delta_T \left(2c - (\epsilon_{T \cup \{i\} }  + \epsilon_T) \right)\right) }{\left( 2V -  \delta_S \left(2c - (\epsilon_{S \cup \{i\} }  + \epsilon_S) \right) \right) }   \\
& \ge  \frac{|T|}{|S|} \frac{(\epsilon_{T \cup \{i\}} - \epsilon_T)}{(\epsilon_{S \cup \{i\}} - \epsilon_S) } \\
& \ge  \frac{(n-1) }{(n-2)} \frac{(\epsilon_{T \cup \{i\}} - \epsilon_T)}{(\epsilon_{S \cup \{i\}} - \epsilon_S) } \\
& \ge 1,
\end{align*}
where the last inequality follows from 
Assumption 1.  When $\epsilon_{S \cup \{i\}} - \epsilon_S = 0$,  we have $G_i(T) \ge G_i(S) = 0$.

We finally show $H_i(T) \ge H_i(S)$.
This can be shown as follows:
\begin{align*}
H_i(T) &= f(\epsilon_{T \cup \{i\}}, 1+ \frac{r (n -2|T| -1) }{n} ) \\
  &  \ge  f(\epsilon_{S \cup \{i\}}, 1+ \frac{r (n -2|T| -1) }{n} )  \\
    &  \ge  f(\epsilon_{S \cup \{i\}}, 1+ \frac{r (n -2|S| -1) }{n} )  \\
    & = H_i(S)
\end{align*}
where the first inequality follows from $\epsilon_{T \cup \{i\}} \ge \epsilon_{S \cup \{i\}}$,
$ 1+ \frac{r (n -2|T| -1) }{n} \le 1+r$, and Lemma \ref{lemma:oligopoly1}(a), 
and  the second inequality follows from  $1+ \frac{r (n -2|S| -1) }{n} \ge 
1+ \frac{r (n -2|T| -1) }{n}$,
$\epsilon_{S \cup \{i\}} \in [0, c]$ and Lemma \ref{lemma:oligopoly1}(b).

(b) 
Fix any $i \in N$ and $S,  T \subseteq N \setminus \{i\}$  such that  $|S| \neq 1$ and $S \subseteq T$. 
Suppose that $|S| \ge 2$.
Then, we have
\begin{align*}
\nu(T \cup \{i\}) -\nu(T) &= v(|T \cup \{i\}|) -v(|T|) + F_i(T) \\
& \ge v(|S \cup \{i\}|) -v(|S|) + F_i(S) \\
& = \nu(S \cup \{i\}) -\nu(S),
\end{align*}
where  the inequality follows from convexity of $v$ and $F_i(T) \ge F_i(S)$.

Suppose that $|S| = 0$ and $|T| = 1$.
Note that $\nu(S') \ge v(S')$ for all $S' \subseteq N$
and $\nu(S') = v(S')$ for all $S' \subseteq N$ with  $|S'| \le 1$ by definition.
Thus,
 \begin{align*}
\nu(T \cup \{i\}) -\nu(T) &\ge v(|T \cup \{i\}|) -v(|T|)  \ge v(|S \cup \{i\}|) -v(|S|)
= \nu( S \cup \{i\}) - \nu(S)
\end{align*}
where the second inequality follows from convexity of $v$.

Suppose that $|S| = 0$ and $|T| \ge 2$. By (\ref{eq:oligopoly1}),
\begin{align*}
\nu(T \cup \{i\}) -\nu(T) = v(|T \cup \{i\}|) -v(|T|) + F_i(T) 
\ge v(|S \cup \{i\}|) -v(|S|) = \nu(S \cup \{i\}) -\nu(S),
\end{align*}
where the inequality follows from convexity of $v$ and $F_i(T) \ge 0$.
\end{proof}

The following lemma simplifies the condition for
3-distant convexity/convexity of $\nu$.

\begin{lemma}\label{lemma:oligopoly3}
\begin{itemize}
\item[(a)] $\nu$ is convex if and only if $\nu(T \cup \{i\} ) - \nu(T)
\ge \nu(S \cup \{i\}) - \nu(S)$
for all $i \in N$ and $S, T \subseteq N \setminus \{i\} $ such that 
$S \subseteq T$ and $|S| = 1$ and $|T| = 2$.
\item[(b)] $\nu$ is $3$-distant convex if and only if $\nu(T \cup \{i\} ) - \nu(T)
\ge \nu(S \cup \{i\}) - \nu(S)$
for all $i \in N$ and $S, T \subseteq N \setminus \{i\} $ such that 
$S \subseteq T$ and $|S| = 1$ and $|T| = 3$.
\end{itemize}
\end{lemma}

\begin{proof}
 For both (a) and (b), the ``only if" parts follow from definition.  Thus, we show the ``if" parts.  Let $i\in N$ and $S,T\subseteq N\setminus \{ i\} $ with $S\subseteq T$. It suffices to consider the case where $S\subsetneq T$.  If $|S|\neq 1$, then the proof is done by Lemma \ref{lemma:oligopoly2}(b).  Thus, assume that $|S|=1$.  

(a)  If $|T|=2$, then the proof follows from the assumption.  Thus, assume that $|T|>2$.  
%
Pick any $j \in T \setminus S $.
Then,
$$
\nu(T \cup \{i\}) - \nu(T) \ge \nu(S \cup \{j\} \cup \{i\}) - \nu(S \cup \{j\})
\ge \nu(S \cup \{i\}) - \nu(S), 
$$
where the first inequality follows from 
Lemma \ref{lemma:oligopoly2}(b).
Thus, $\nu$ is convex.

(b)  It suffices to consider the case where $|T|\ge 3$ to show that $\nu $ is $3$-distant convex. If $|T|=3$, then the proof follows from the assumption.  Thus, assume that $|T|>3$.    
Take any distinct $j_1, j_2 \in T \setminus S$.
Then,
$$
\nu(T \cup \{i\}) - \nu(T) \ge \nu(S \cup \{j_1, j_2\} \cup \{i\}) - \nu(S \cup \{j_1, j_2\})
\ge \nu(S \cup \{i\}) - \nu(S),
$$
where the first inequality follows from 
Lemma \ref{lemma:oligopoly2}(b).
Thus, $\nu$ is $3$-distant convex by Proposition \ref{prop:marginal}.
\end{proof}

We now show Proposition \ref{prop:symmetric_oligopoly}.
We assume that  $\epsilon_S = \epsilon \ge 0$ for all $S \subseteq N$
with $|S| \ge 2$. 
 Note that $G_i(S)=0$ for all $S\subseteq N$ in this case. Since $\nu$ depends only on the coalition size, we write $\nu(s)$ for any coalition of size $s$.
Fix any $k = 2, 3$.
We show that $\nu$ is $k$-distant convex if and only if $\epsilon \le \epsilon_k$,
where $2$-distant convex means the usual convexity.
By Lemma \ref{lemma:oligopoly3},
 it suffices to show that $\nu (k+1) - \nu(k) \ge \nu(2) - \nu(1)$  if and only if $\epsilon \le \epsilon_k$.
By (\ref{eq:oligopoly1}),
$$
\nu(k+1) - \nu(k) = v(k+1) - v(k) - \frac{1}{4}  (2c\epsilon - \epsilon^2 ) \left(1 + \frac{ r (n-2k -1) }{n} \right) + \frac{\epsilon V}{2}.
$$
By (\ref{eq:oligopoly0}) and $\nu(1) = v(1)$,
$$
\nu(2) - \nu(1) = v(2) - v(1) - \frac{1}{2} (2c\epsilon  - \epsilon^2 )\left(1 +  \frac{r(n - 2)}{n} \right) + \epsilon V
$$
Let
$$
f_k(\epsilon) = \nu(k+1) - \nu(k) - (\nu(2) - \nu(1)) = - \frac{1}{4}L_k \epsilon^2 +\frac{1}{2} (cL_k -V) \epsilon + M_k,
$$
where $L_k = 1+ \frac{r(n-3+2k)}{n} > 0$ and $M_k = v(k+1) -v(k) - (v(2)- v(1)) > 0$.
Then,
$\nu (k+1) - \nu(k) \ge \nu(2) - \nu(1)$ if and only if $ f_k(\epsilon) \ge 0$.
The following quadratic equation
$$
-L_k \epsilon^2 + 2 (cL_k -V) \epsilon + 4M_k = 0
$$
has the two solutions given by
$$
\epsilon_k = \frac{(c L_k -V) + \sqrt{(c L_k -V)^2 + 4 L_k M_k}    }{L_k} > 0 \text{ and } 
\frac{(c L_k -V) - \sqrt{(c L_k -V)^2 + 4 L_k M_k}    }{L_k} < 0.
$$
By $\epsilon \ge 0$, $\nu(k+1) - \nu(k) - (\nu(2) - \nu(1)) \ge 0$ if and only if $\epsilon \le \epsilon_k$.
Clearly, $\epsilon_k > 0$.   Thus, (b) and (c) are proved. 
By strict convexity of $v$ and Lemma \ref{lemma:oligopoly2}(a),
$$
\nu(4) - \nu(3) = v(4) -v(3) +  G_i(3) >  v(3) -v(2) + G_i(2) = \nu(3)- \nu(2).
$$
 Thus, $f_3(\epsilon) > f_2(\epsilon)$. This implies $\epsilon_3 > \epsilon_2$, completing the proof of (a).

We finally show Proposition \ref{prop:asymmetry_oligopoly}.
Let $k = 2, 3$.
Let $i \in N$ and $S, T \subseteq N \setminus \{i\}$
such that $|S| = 1$ and $|T| = k$.
We denote $S = \{ j\}$.
By Lemma \ref{lemma:oligopoly3},
it suffices to show $\nu(T \cup \{i\}) - \nu(T) \ge
\nu(\{j, i\}) - \nu(\{j\})$.
By (\ref{eq:oligopoly1}) and the definition of $f$,
we have
\begin{align}
\nu(T \cup \{i\}) -\nu(T) & = v(k+1) - v(k)  + 
G_i(T) + H_i(T) \notag \\
& \ge v(k+1) - v(k) + H_i(T)\notag \\
&  =  v(k+1) - v(k)  +f(\epsilon_{T \cup\{i\}}, 1 +  \frac{ r (n - 2k -1) }{n}  ) \notag \\
& \ge v(k+1) - v(k)  +f(\epsilon_{\{j, i\}}, 1 +  \frac{ r (n - 2k -1)  }{n}  ) \notag \\
& =  v(k+1) - v(k) - \frac{1}{4} \epsilon_{\{j, i\}} (2c - \epsilon_{\{ j, i\}} ) \left(1 + \frac{ r (n-2k -1) }{n} \right) + \frac{\epsilon_{\{j, i\}} V}{2} \label{eq:oligopoly3},
\end{align}
 where the first inequality follows from $G_i(T) \ge 0$
 and the second inequality follows from Lemma \ref{lemma:oligopoly1} (b). 
By \eqref{eq:oligopoly0} and $v(1)=\nu (1)$, we
also have
\begin{align}
\nu(\{j, i\}) - \nu(\{j\}) = v(2) -v(1) - 
\frac{1}{2} ( 2 \epsilon_{ \{j, i\} } c - \epsilon_{\{j,i \} }^2 ) \left(1 + r \frac{(n - 2)}{n}\right) +  \epsilon_{ \{j, i\} } V. \label{eq:oligopoly4}
\end{align}
Therefore, $\nu(T \cup \{i\}) -\nu(T) -  (\nu(\{j, i\}) - \nu(\{j\}))\ge 0$ is implied by 
$(\ref{eq:oligopoly3}) - (\ref{eq:oligopoly4}) \ge 0$.
Note that $(\ref{eq:oligopoly3}) - (\ref{eq:oligopoly4}) \ge 0$ if and only if
 $f_k(\epsilon_{ \{i, j\}}) \ge 0$.
By the proof of Proposition \ref{prop:symmetric_oligopoly},
 $\epsilon_{ \{ i, j\} } \le \epsilon_k$ if and only if $f_k(\epsilon_{ \{i, j\}}) \ge 0$.
Therefore, this completes the proof of Proposition \ref{prop:asymmetry_oligopoly}.

\subsection*{Proof of the nonemptiness of the core in Section \ref{subsec:oligopoly}}

Let $v'$ be the $\alpha$-characteristic function in the case
when each firm has the identical marginal cost $c_N =  c -\epsilon_N$ and there are no synergy effects.
As $v'$ is convex by the result of Lardon, the equal split allocation $x$, defined by $x_i = \frac{v'(N)}{n}$ for all $i \in N$, is in the core of $(N, v')$.
Note that $v'(N) = \nu(N)$.
Moreover, $v'(S)\ge \nu(S)$ as
$ c_{S} \ge c_N$ for all $S \subseteq N$.
This implies that $x$ is also in the core of $(N, \nu)$.

\subsection*{Proof of Proposition \ref{prop:production}}

Fix arbitrary $S,T \subseteq N$ such that $S\setminus T\neq \emptyset $, $T\setminus S\neq \emptyset $, and 
$|S\triangle T|>2$.  Without loss of generality, we may assume that $w(f^2(S))\le w(f^2(T))$.  
Let $r<1$ be a real number that is sufficiently close to $1$.  
Then, it suffices to show that 
$w(f^1(S))w(f^2(S))+w(f^1(T))w(f^2(T))<w(f^1(S\cap T))w(f^2(S\cap T))+w(f^1(S\cup T))w(f^2(S\cup T))$.  
To this end, we introduce the following basic properties.

\begin{claim}\label{app_basic}
Let $S \subseteq N$.  Then, $w(S')w(S\setminus S')<w(S'')w(S\setminus S'')$ if and only if $\min \{ w(S'),w(S\setminus S')\} <\min \{ w(S''),w(S\setminus S'')\} $. 
\end{claim}
\begin{proof} 
It immediately follows from $w(S')w(S\setminus S')=w(S')(w(S)-w(S'))$ for any $S'\subseteq S$ and the property of the quadratic function.  
\end{proof}



\begin{claim}\label{both}
Let $ S,T \subseteq N$ such that $S\setminus T\neq \emptyset $, $T\setminus S\neq \emptyset $, and $w(f^2(S))\le w(f^2(T))$.  
Then, $\min _{p=1,2}w(f^p(S)\cap T)\le w(f^1(T))$ and $\max _{p=1,2}w(f^p(S)\cap T)\le w(f^2(T))$, where at least one of these inequalities holds strictly.    
\end{claim}
\begin{proof}
It follows that $\max _{p=1,2}w(f^p(S)\cap T)\le w(f^2(T))$ from $w(f^2(S))\le w(f^2(T))$.  
Suppose that $\min _{p=1,2}w(f^p(S)\cap T)>w(f^1(T))$.  Then, 
\begin{align}\label{half}
\frac{w(T)}{2}>\frac{w(S\cap T)}{2}\ge \min _{p=1,2}w(f^p(S)\cap T)>w(f^1(T)), 
\end{align}
where the first inequality follows from $T\setminus S\neq \emptyset $ and the second inequality follows from $(f^1(S)\cap T)\cup (f^2(S)\cap T)=S\cap T$.  Denote $a=\min _{p=1,2}w(f^p(S)\cap T)$.  Note that $f^p(S)\cap T\subseteq T$ for each $p=1,2$.  
Then, \eqref{half} and Claim \ref{app_basic} imply $w(f^1(T))w(f^2(T))<a(w(T)-a)$, contradicting the definition of $f^1(T)$.  Hence, $\min _{p=1,2}w(f^p(S)\cap T)\le w(f^1(T))$. 
Moreover, since $w(T)>w(S\cap T)$ by $T\setminus S\neq \emptyset $, we have 
$\min _{p=1,2}w(f^p(S)\cap T)<w(f^1(T))$ or  $\max _{p=1,2}w(f^p(S)\cap T)< w(f^2(T))$. 
\end{proof}

The remaining proof is done by distinguishing two cases.  
\bigskip

\noindent {\bf Case 1.} $f^p(S)\setminus T\neq \emptyset $ for each $p=1,2$.  

We have 
\begin{align}
A\cap B&=\emptyset \text{ for any distinct }A,B\in \{ f^1(T),f^2(T),f^1(S)\setminus T,f^2(S)\setminus T\} \label{set0a} \\
f^1(T)&\cup f^2(T)\cup \left( f^1(S)\setminus T\right)\cup \left(f^2(S)\setminus T\right)=S\cup T. \label{set0b}
\end{align}
Then, 
\begin{align*}
w&(f^1(S))w(f^2(S))+w(f^1(T))w(f^2(T))\\ 
&=\left(w(f^1(S)\cap T)+w(f^1(S)\setminus T)\right)\left(w(f^2(S)\cap T)+w(f^2(S)\setminus T)\right)+w(f^1(T))w(f^2(T)) \\ 
&=w(f^1(S)\cap T)w(f^2(S)\cap T)+\underline{ w(f^1(S)\cap T)}w(f^2(S)\setminus T)+w(f^1(S)\setminus T)\underline{w(f^2(S)\cap T)} \\ 
&\ \ \ \ \ +w(f^1(S)\setminus T)w(f^2(S)\setminus T)+w(f^1(T))w(f^2(T)) \\ 
&<w(f^1(S)\cap T)w(f^2(S)\cap T)+\underline{w(f^1(T))}w(f^2(S)\setminus T)+w(f^1(S)\setminus T)\underline{ w(f^2(T)) }\\ 
&\ \ \ \ \ +w(f^1(S)\setminus T)w(f^2(S)\setminus T)+w(f^1(T))w(f^2(T)) \\ 
&=w(f^1(S)\cap T)w(f^2(S)\cap T)+\left(w(f^1(T))+w(f^1(S)\setminus T)\right)\left(w(f^2(T))+w(f^2(S)\setminus T)\right) \\ 
&\le w(f^1(S\cap T))w(f^2(S\cap T))+w(f^1(S\cup T))w(f^2(S\cup T)), 
\end{align*}
where the strict inequality follows from Claim \ref{both} and the assumptions that $w(f^2(S))\le w(f^2(T))$ and $f^p(S)\setminus T\neq \emptyset $ for each $p=1,2$ (the underlined terms highlight the terms that differ on the two sides of the strict inequality), and the weak inequality follows from  $(f^1(S)\cap T)\cap (f^2(S)\cap T)=\emptyset $, $(f^1(S)\cap T)\cup (f^2(S)\cap T)=S\cap T$, \eqref{set0a}, and \eqref{set0b}.  

\bigskip 

\noindent {\bf Case 2.} $f^p(S)\setminus T=\emptyset $ for some $p=1,2$. 

Note that $f^p(S)\setminus T=\emptyset $ holds for exactly one of 1 and 2 by $S\setminus T\neq \emptyset $.  
Note also that $S\setminus T\subseteq f^1(S)$ or $S\setminus T\subseteq f^2(S)$ holds in this case.  

Let $q=1,2$ denote the index such that $f^q(S)\setminus T\neq \emptyset $ and 
$q'=1,2$ with $q'\neq q$ denote the index such that $f^{q'}(S)\setminus T=\emptyset $.  Note that $f^{q}(S)\setminus T=S\setminus T$.  
By $f^{q'}(S)\subseteq T$, we have 
\begin{align}
(f^q(S)\cap T)\cap f^{q'}(S)&=\emptyset \label{set1a}; \\
(f^q(S)\setminus T)\cap f^p(T)&=\emptyset \text{ for each }p=1,2, \label{set1b}\\
(f^q(S)\cap T)\cup f^{q'}(S)&=S\cap T, \label{set1c} \\
(f^q(S)\setminus T)\cup f^1(T)\cup f^2(T)&=S\cup T.\label{set1d}
\end{align}  

\noindent {\bf Subcase 2(a)} $w(f^2(T))>w(f^2(S))$. 

Note that we also have $w(f^2(T))>w(f^1(S))$.  
Then, 
\begin{align*}
w&(f^1(S))w(f^{2}(S))+w(f^1(T))w(f^2(T)) \\ 
&=\left(w(f^q(S)\cap T)+w(f^q(S)\setminus T)\right)w(f^{q'}(S))+w(f^1(T))w(f^2(T)) \\ 
&=w(f^q(S)\cap T)w(f^{q'}(S))+w(f^q(S)\setminus T)\underline{ w(f^{q'}(S))}+w(f^1(T))w(f^2(T)) \\ 
&<w(f^q(S)\cap T)w(f^{q'}(S))+w(f^q(S)\setminus T)\underline{ w(f^{2}(T))}+w(f^1(T))w(f^2(T)) \\ 
&=w(f^q(S)\cap T)w(f^{q'}(S))+\left(w(f^q(S)\setminus T)+w(f^1(T))\right)w(f^2(T)) \\ 
&\le w(f^1(S\cap T))w(f^2(S\cap T))+w(f^1(S\cup T))w(f^2(S\cup T)), 
\end{align*}
where the strict inequality follows from $w(f^2(T))>w(f^2(S))\ge w(f^1(S))$(the underlined terms highlight the terms that differ on the two sides of the strict inequality), and the weak inequality follows from $f^1(T)\cap f^2(T)=\emptyset $, \eqref{set1a}, and \eqref{set1b}.   

\bigskip 

\noindent {\bf Subcase 2(b)} $w(f^2(T))=w(f^2(S))$.  

Since $|S\triangle T|>2$, we may choose $S$ and $T$ to satisfy $|S\setminus T|\ge 2$ in this case without any loss of generality.  

We begin with the subsubcase where $w(f^2(S))>w(f^1(S))$ and $f^1(S)\setminus T=\emptyset $.  
Note that $f^2(S)\setminus T=S\setminus T\neq \emptyset $.  This case can be shown in a similar way to the previous case:   
\begin{align*}
w&(f^1(S))w(f^{2}(S))+w(f^1(T))w(f^2(T)) \\ 
&=w(f^1(S))\left(w(f^2(S)\cap T)+w(f^2(S)\setminus T)\right)+w(f^1(T))w(f^2(T)) \\ 
&=w(f^1(S))w(f^2(S)\cap T)+\underline{ w(f^1(S))}w(f^{2}(S)\setminus T)+w(f^1(T))w(f^2(T)) \\ 
&<w(f^1(S))w(f^2(S)\cap T)+\underline{ w(f^2(T))}w(f^{2}(S)\setminus T)+w(f^1(T))w(f^2(T)) \\ 
&=w(f^1(S))w(f^2(S)\cap T)+w(f^2(T))\left(w(f^{2}(S)\setminus T)+w(f^1(T))\right) \\ 
&\le w(f^1(S\cap T))w(f^2(S\cap T))+w(f^1(S\cup T))w(f^2(S\cup T)), 
\end{align*}
where the strict inequality follows from $w(f^2(T))=w(f^2(S))>w(f^1(S))$ and $f^2(S)\setminus T\neq \emptyset $ (the underlined terms highlight the terms that differ on the two sides of the strict inequality), and the weak inequality follows from $f^1(T)\cap f^2(T)=\emptyset $, \eqref{set1a}-\eqref{set1d}.   

We turn to the subsubcase where $w(f^2(S))=w(f^1(S))$ or $f^2(S)\setminus T=\emptyset $.   
Since $f^2(S)\setminus T=\emptyset $ if and only if $f^1(S)\setminus T\neq \emptyset $ in this case,  this condition is equivalent to that 
$w(f^2(S))=w(f^1(S))$ or $f^1(S)\setminus T\neq \emptyset $.  
This condition is also equivalent to that $f^1(S)\subseteq T$ implies $w(f^2(S))=w(f^1(S))$.  
Thus, when $f^1(S)\subseteq T$ happens,  we can exchange $f^1(S)$ and $f^2(S)$ since $f^1(S),f^2(S)\in \text{arg}\max _{S'\subseteq S}w(S')(w(S)-w(S'))$ and we only require that $w(f^1(S))\le w(f^2(S))$.  
Therefore, we may assume that $f^2(S)\subseteq T$ because either $f^1(S)\subseteq T$ or $f^2(S)\subseteq T$ holds in this case.  This implies that $S\setminus T\subseteq f^1(S)$ in either case.  

Then, 
\begin{align*}
w&(f^1(S))w(f^{2}(S))+w(f^1(T))w(f^2(T)) \\ 
&=\left(w(f^1(S)\cap T)+w(f^1(S)\setminus T)\right)w(f^2(S))+w(f^1(T))w(f^2(T)) \\ 
&=w(f^1(S)\cap T)w(f^2(S))+w(f^{1}(S)\setminus T)\underline{w(f^2(S))}+w(f^1(T))w(f^2(T)) \\ 
&=w(f^1(S)\cap T)w(f^2(S))+w(f^{1}(S)\setminus T)\underline{w(f^2(T))}+w(f^1(T))w(f^2(T)) \\ 
&=w(f^1(S)\cap T)w(f^2(S))+\left(w(f^{1}(S)\setminus T)+w(f^1(T))\right)w(f^2(T)) 
\end{align*}
where the third equality follows from $w(f^2(S))=w(f^2(T))$ (the underlined terms highlight the terms that differ on the two sides of the fourth equality).  

By \eqref{set1b}, \eqref{set1d}, and $f^1(T)\cap f^2(T)=\emptyset $, $(w(f^{1}(S)\setminus T)+w(f^1(T)))w(f^2(T))\le w(f^1(S\cup T))w(f^2(S\cup T))$.  
Therefore, the proof is done if  we show that $w(f^1(S)\cap T)w(f^2(S))<w(f^1(S\cap T))w(f^2(S\cap T))$. 
Fix an arbitrary $k\in f^2(S)$.  By $f^2(S)\subseteq T$, we have $S\setminus T\subseteq f^1(S)$.  
Therefore, 
$S\setminus T=f^1(S)\setminus T$.   Since $|f^1(S)\setminus T|=|S\setminus T|\ge 2$, 
$w(f^1(S)\setminus T)>w _k$.  Then, 
\begin{align*}
w(f^2(S)\setminus \{ k\} )=w(f^2(S))-w_k>w(f^1(S))-w(f^1(S)\setminus T)=w(f^1(S)\cap T).  
\end{align*}
Thus, $w(f^1(S)\cap T)=\min \{ w(f^1(S)\cap T),w(f^2(S))\} $.  It also follows that $w(f^2(S)\setminus \{ k\} )>w(f^1(S)\cap T)$ from this inequality.  Moreover, $w((f^1(S)\cap T)\cup \{ k\} )>w(f^1(S)\cap T)$ obviously holds.    By Claim \ref{app_basic}, $w(f^1(S)\cap T)w(f^2(S))<w((f^1(S)\cap T)\cup \{ k\} )w(f^2(S)\setminus \{ k\} )$.  
By \eqref{set1a}, and \eqref{set1c},  $((f^1(S)\cap T)\cup \{ k\} )\cap (f^2(S)\setminus \{ k\} )=\emptyset $ and $((f^1(S)\cap T)\cup \{ k\} )\cup (f^2(S)\setminus \{ k\} )=S\cap T$.  Then, 
$w((f^1(S)\cap T)\cup \{ k\} )w(f^2(S)\setminus \{ k\} )\le w(f^1(S\cap T))w(f^2(S\cap T))$.  
Hence, $w(f^1(S)\cap T)w(f^2(S))<w(f^1(S\cap T))w(f^2(S\cap T))$, the desired inequality.


\bibliographystyle{chicago}
\bibliography{reference}

@book{Schrijver2003,
author    = {Schrijver, Alexander},
title     = {Combinatorial Optimization: Polyhedra and Efficiency},
series    = {Algorithms and Combinatorics},
volume    = {24},
publisher = {Springer},
address   = {Berlin, Heidelberg},
year      = {2003},
isbn      = {978-3-540-44389-6}
}

@book{Frank2011,
  author    = {Frank, Andr{\'a}s},
  title     = {Connections in Combinatorial Optimization},
  series    = {Oxford Lecture Series in Mathematics and Its Applications},
  volume    = {38},
  publisher = {Oxford University Press},
  address   = {Oxford},
  year      = {2011},
  isbn      = {9780199205271}
}

@article{Williamson1968,
author  = {Williamson, Oliver E.},
title   = {Economies as an Antitrust Defense: The Welfare Tradeoffs},
journal = {The American Economic Review},
year    = {1968},
volume  = {58},
number  = {1},
pages   = {18--36}
}

@article{FarrellShapiro1990,
  author  = {Farrell, Joseph and Shapiro, Carl},
  title   = {Horizontal Mergers: An Equilibrium Analysis},
  journal = {American Economic Review},
  year    = {1990},
  volume  = {80},
  number  = {1},
  pages   = {107--126}
}

@article{DeneckereDavidson1985,
author  = {Deneckere, Raymond and Davidson, Carl},
title   = {Incentives to Form Coalitions with Bertrand Competition},
journal = {The RAND Journal of Economics},
year    = {1985},
volume  = {16},
number  = {4},
pages   = {473--486},
doi     = {10.2307/2555507}
}

@article{Moulin1985,
  author  = {Moulin, Herv{\'e}},
  title   = {The Separability Axiom and Equal-Sharing Methods},
  journal = {Journal of Economic Theory},
  year    = {1985},
  volume  = {36},
  number  = {1},
  pages   = {120--148},
  doi     = {10.1016/0022-0531(85)90082-1}
}

@book{Shubik1980MarketStructure,
  title     = {Market Structure and Behavior},
  author    = {Shubik, Martin},
  year      = {1980},
  publisher = {Harvard University Press},
  address   = {Cambridge, MA}
}

@article{HouLardonDriessen2025Bertrand,
  title   = {Convexity and the {Shapley} Value of {Bertrand} Oligopoly TU-Games in $\beta$-Characteristic Function Form},
  author  = {Hou, Dongshuang and Lardon, Aymeric and Driessen, Theo S. H.},
  journal = {Theory and Decision},
  volume  = {98},
  pages   = {519--536},
  year    = {2025},
  publisher = {Springer},
  doi     = {10.1007/s11238-024-10022-y}
}

@article{DriessenMeinhardt2001,
  author  = {Driessen, Theo S. H. and Meinhardt, Holger Ingmar},
  title   = {{(Average-)} Convexity of Common Pool and Oligopoly {TU}-Games},
  journal = {International Game Theory Review},
  year    = {2001},
  volume  = {3},
  number  = {2--3},
  pages   = {141--158},
  doi     = {10.1142/S0219198901000336}
}

@article{Meinhardt1999CommonPool,
  title   = {Common Pool Games Are Convex Games},
  author  = {Meinhardt, Holger},
  journal = {Journal of Public Economic Theory},
  volume  = {1},
  number  = {2},
  pages   = {247--270},
  year    = {1999},
  publisher = {Blackwell}
}

@article{Zhao1999ConvexityOligopoly,
  title   = {A Necessary and Sufficient Condition for the Convexity in Oligopoly Games},
  author  = {Zhao, Jingang},
  journal = {Mathematical Social Sciences},
  volume  = {37},
  number  = {2},
  pages   = {189--204},
  year    = {1999},
  publisher = {Elsevier},
  doi     = {10.1016/S0165-4896(98)00019-5}
}

@article{CurielPederzoliTijs1989Sequencing,
  title   = {Sequencing Games},
  author  = {Curiel, Imma and Pederzoli, Giorgio and Tijs, Stef},
  journal = {European Journal of Operational Research},
  volume  = {40},
  number  = {3},
  pages   = {344--351},
  year    = {1989},
  publisher = {Elsevier},
  doi     = {10.1016/0377-2217(89)90276-8}
}

@incollection{CurielHamersKlijn2002Survey,
  title     = {Sequencing Games: A Survey},
  author    = {Curiel, Imma and Hamers, Herbert and Klijn, Flip},
  booktitle = {Chapters in Game Theory},
  editor    = {Borm, Peter and Peters, Hans},
  pages     = {27--50},
  year      = {2002},
  publisher = {Kluwer Academic Publishers},
  address   = {Dordrecht}
}

@article{littlechild_owen1973,
  title   = {A Simple Expression for the {Shapley} Value in a Special Case},
  author  = {Littlechild, Stephen C. and Owen, Guillermo},
  journal = {Management Science},
  volume  = {20},
  number  = {3},
  pages   = {370--372},
  year    = {1973},
  publisher = {INFORMS}
}

@article{Lardon2019Core,
  author    = {Aymeric Lardon},
  title     = {The Core in {Bertrand} Oligopoly {TU}-Games with Transferable Technologies},
  journal   = {The B.E. Journal of Theoretical Economics},
  year      = {2019},
  volume    = {},
  number    = {},
  pages     = {20180197},
  doi       = {10.1515/bejte-2018-0197},
  publisher = {De Gruyter}
}

@article{Lardon2020Convexity,
  author    = {Aymeric Lardon},
  title     = {Convexity of {Bertrand} oligopoly {TU}-games with differentiated products},
  journal   = {Annals of Operations Research},
  year      = {2020},
  volume    = {287},
  number    = {},
  pages     = {285--302},
  doi       = {10.1007/s10479-019-03351-7},
  publisher = {Springer}
}

@article{Bahel2026Preconvex,
  title   = {Preconvex Games},
  author  = {Bahel, Eric and Trudeau, Christian and Wang, Haoyu},
  journal = {Games and Economic Behavior},
  volume  = {155},
  pages   = {250--266},
  year    = {2026},
  publisher = {Elsevier},
  doi     = {10.1016/j.geb.2025.11.001}
}

@article{Leng2021ADMC,
  title   = {Multiplayer Allocations in the Presence of Diminishing Marginal Contributions: Cooperative Game Analysis and Applications in Management Science},
  author  = {Leng, Mingming and Luo, Chunlin and Liang, Liping},
  journal = {Management Science},
  volume  = {67},
  number  = {5},
  pages   = {2891--2903},
  year    = {2021},
  publisher = {INFORMS},
  doi     = {10.1287/mnsc.2020.3709}
}

@article{DriessenTijs1985Semiconvex,
  title   = {The r-Value, the Core and Semiconvex Games},
  author  = {Driessen, Theo S. H. and Tijs, Stef H.},
  journal = {International Journal of Game Theory},
  volume  = {14},
  number  = {4},
  pages   = {229--248},
  year    = {1985},
  publisher = {Physica-Verlag}
}

@article{Bahel2021Hyperadditive,
  title   = {Hyperadditive Games and Applications to Networks or Matching Problems},
  author  = {Bahel, Eric},
  journal = {Journal of Economic Theory},
  volume  = {191},
  pages   = {105168},
  year    = {2021},
  publisher = {Elsevier},
  doi     = {10.1016/j.jet.2020.105168}
}

@article{Sprumont1990PMAS,
  title   = {Population Monotonic Allocation Schemes for Cooperative Games with Transferable Utility},
  author  = {Sprumont, Yves},
  journal = {Games and Economic Behavior},
  volume  = {2},
  number  = {4},
  pages   = {378--394},
  year    = {1990},
  publisher = {Academic Press}
}

@article{Driessen1986KConvex,
  title   = {Solution Concepts of k-Convex n-Person Games},
  author  = {Driessen, Theo S. H.},
  journal = {International Journal of Game Theory},
  volume  = {15},
  number  = {4},
  pages   = {201--229},
  year    = {1986},
  publisher = {Physica-Verlag}
}

@article{DengIbarakiNagamochi1999Core,
  title   = {Algorithmic Aspects of the Core of Combinatorial Optimization Games},
  author  = {Deng, Xiaotie and Ibaraki, Toshihide and Nagamochi, Hiroshi},
  journal = {Mathematics of Operations Research},
  volume  = {24},
  number  = {3},
  pages   = {751--766},
  year    = {1999},
  publisher = {INFORMS}
}

@article{DengPapadimitriou1994Complexity,
  title   = {On the Complexity of Cooperative Solution Concepts},
  author  = {Deng, Xiaotie and Papadimitriou, Christos H.},
  journal = {Mathematics of Operations Research},
  volume  = {19},
  number  = {2},
  pages   = {257--266},
  year    = {1994},
  publisher = {INFORMS}
}

@article{ONeill1982,
  title   = {A Problem of Rights Arbitration from the Talmud},
  author  = {O'Neill, Barry},
  journal = {Mathematical Social Sciences},
  volume  = {2},
  number  = {4},
  pages   = {345--371},
  year    = {1982},
  doi     = {10.1016/0165-4896(82)90029-4}
}

@article{CurielMaschlerTijs1988Bankruptcy,
  title   = {Bankruptcy Games},
  author  = {Curiel, Ignacio and Maschler, Michael and Tijs, Stef},
  journal = {Zeitschrift f{\"u}r Operations Research},
  volume  = {31},
  pages   = {A143--A159},
  year    = {1988},
  doi     = {10.1007/BF02189458}
}

@article{AumannMaschler1985BankruptcyTalmud,
  title   = {Game Theoretic Analysis of a Bankruptcy Problem from the Talmud},
  author  = {Aumann, Robert J. and Maschler, Michael},
  journal = {Journal of Economic Theory},
  volume  = {36},
  number  = {2},
  pages   = {195--213},
  year    = {1985},
  doi     = {10.1016/0022-0531(85)90055-5}
}

@book{PelegSudholter2007CooperativeGames,
  title     = {Introduction to the Theory of Cooperative Games},
  author    = {Peleg, Bezalel and Sudh{\"o}lter, Peter},
  edition   = {2},
  year      = {2007},
  publisher = {Springer},
  address   = {Berlin, Heidelberg},
  series    = {Theory and Decision Library C: Game Theory, Mathematical Programming and Operations Research},
  volume    = {34},
  isbn      = {978-3-540-72944-8},
  doi       = {10.1007/978-3-540-72945-5}
}

@article{Ichiishi1981,
  author  = {Ichiishi, Tatsuro},
  title   = {Super-modularity: Applications to Convex Games and to the Greedy Algorithm for {LP}},
  journal = {Journal of Economic Theory},
  volume  = {25},
  year    = {1981},
  pages   = {283--286}
}

@incollection{Gillies1959,
  author    = {Gillies, Donald B.},
  title     = {Solutions to General Non-Zero-Sum Games},
  booktitle = {Contributions to the Theory of Games IV},
  editor    = {Tucker, Albert W. and Luce, R. Duncan},
  publisher = {Princeton University Press},
  year      = {1959},
  pages     = {47--85},
  address   = {Princeton, NJ}
}

@article{Bondareva1963,
  author  = {Bondareva, Olga N.},
  title   = {Some Applications of Linear Programming Methods to the Theory of Cooperative Games},
  journal = {Problemy Kibernetiki},
  volume  = {10},
  year    = {1963},
  pages   = {119--139},
  note    = {In Russian}
}

@article{Shapley1967,
  author  = {Shapley, Lloyd S.},
  title   = {On Balanced Sets and Cores},
  journal = {Naval Research Logistics Quarterly},
  volume  = {14},
  year    = {1967},
  pages   = {453--460}
}

@article{Granot1982Perm,
  title = {The Relationship Between Convex Games and Minimum Cost Spanning Tree Games: A Case for Permutationally Convex Games},
  volume = {3},
  number = {3},
  journal = {SIAM Journal on Algebraic Discrete Methods},
  author = {Granot,  Daniel and Huberman,  Gur},
  year = {1982},
  pages = {288--292}
}

@article{Shapley1971Cores,
  author    = {Lloyd S. Shapley},
  title     = {Cores of Convex Games},
  journal   = {International Journal of Game Theory},
  year      = {1971},
  volume    = {1},
  number    = {1},
  pages     = {11--26},
}

@article{littlechild1974simple,
  title={A simple expression for the nucleolus in a special case},
  author={Littlechild, Stephen C},
  journal={International Journal of Game Theory},
  volume={3},
  number={1},
  pages={21--29},
  year={1974},
  publisher={Springer}
}

@article{Mizutani2024KDist,
  title     = {A Polynomial Algorithm for Minimizing $k$-Distant Submodular Functions},
  author    = {Ryuhei Mizutani},
  year      = {2024},
  journal = {arXiv e-prints},
  pages = {arXiv:2407.05127},
  journal   = {Preprint},
}

@article{Nunez1998almostconvex,
  author = {N{\'u}{\~n}ez, Marina and Rafels, Carles},
  title = {On extreme points of the core and reduced games},
  journal = {Annals of Operations Research},
  year = {1998},
  volume = {84},
  pages = {121--133},
  doi = {https://doi.org/10.1023/A:1018980602195}
}

@article{IU1993IJGT,
	author = {I\~{n}arra, Elena and Usategui, Jos\`{e} M.},
	journal = {International Journal of Game Theory},
    volume = {22},
    number = {1},
    pages  = {13--29},
	title = {The {Shapley} value and average convex games},
	year = {1993}}
\end{document}